\documentclass[12pt,preprint]{aastex}	

\newcommand {\kms}	{{\mbox{$\rm \,km\,s^{-1}$}}}

\newcommand {\msun}	{{\mbox{$\rm \,M_\odot$}}}

\newcommand {\hi}	{{\ion{H}{1}}}

\newcommand {\twco}	{{\mbox{$^{12}$CO}}}

\newcommand {\subhisa}	{{_{_{\!H\!I\!S\!A}}}}

\newcommand {\subon}	{{_{_{\!O\!N}}}}
\newcommand {\suboff}	{{_{_{\!O\!F\!F}}}}
\newcommand {\subunabs}	{{_{_{\!U}}}}
\newcommand {\subcont}	{{_{_{\!C}}}}
\newcommand {\subspin}	{{_{_{\!S}}}}

\newcommand {\rms}      {{\it rms}}
\newcommand {\subrms}	{{_{\!rms}}}

\newcommand {\sublsr}	{{_{_{\!L\!S\!R}}}}

\newcommand {\glon}	{{\ell}}
\newcommand {\glat}	{{b}}
\newcommand {\vlsr}	{{v\sublsr}}
\newcommand {\vel}	{{v}}

\newcommand {\dec}	{{\delta}}

\newcommand {\ton}	{{T\subon}}
\newcommand {\toff}	{{T\suboff}}
\newcommand {\tu}	{{T\subunabs}}
\newcommand {\tc}	{{T\subcont}}
\newcommand {\ts}	{{T\subspin}}

\newcommand {\trms}	{{T\subrms}}

\newcommand {\puckdt}	{{\Delta T_{p}}}
\newcommand {\puckdv}	{{\Delta v_{p}}}
\newcommand {\puckda}	{{\Delta \theta_{p}}}

\newcommand {\epuckdt}	{{\Delta T_{ec}}}
\newcommand {\epuckdv}	{{\Delta v_{ec}}}
\newcommand {\epuckda}	{{\Delta \theta_{ec}}}
\newcommand {\epuckvol}	{{V_{ec}}}
\newcommand {\epuckvolmax}{{V_{ec,max}}}
\newcommand {\dt}	{{\Delta T}}
\newcommand {\dv}	{{\Delta v}}
\newcommand {\da}	{{\Delta \theta}}
\newcommand {\aoff}	{{\theta_{\rm off}}}

\newcommand {\bfrac}	{{p}}
\newcommand {\hifrac}	{{f_n}}

\newcommand {\coldens}	{{N\subhisa}}
\newcommand {\voldens}	{{n\subhisa}}
\newcommand {\mass}	{{M\subhisa}}
\newcommand {\nedge}	{{N\suboff}}

\newcommand {\wnon}	{{w_n}}
\newcommand {\wmos}	{{w_m}}
\newcommand {\via}	{{N_{in,all}}}
\newcommand {\voa}	{{N_{out,all}}}
\newcommand {\vid}	{{N_{in,det}}}
\newcommand {\vot}	{{N_{out,true}}}
\newcommand {\fid}	{{f_{det}}}
\newcommand {\fot}	{{f_{true}}}

\newcommand {\toffavg}	{{\langle \toff \rangle}}

\newcommand {\dti}	{{\dt_{in}}}
\newcommand {\tui}	{{\tu_{in}}}
\newcommand {\dai}	{{\da_{in}}}
\newcommand {\dvi}	{{\dv_{in}}}
\newcommand {\dto}	{{\dt_{out}}}
\newcommand {\tuo}	{{\tu_{out}}}
\newcommand {\dao}	{{\da_{out}}}
\newcommand {\dvo}	{{\dv_{out}}}
\newcommand {\ddt}      {{\Delta\dt}}
\newcommand {\dtu}	{{\Delta\tu}}
\newcommand {\dda}	{{\Delta\da}}
\newcommand {\ddv}	{{\Delta\dv}}

\shorttitle{An Automated HISA Survey Method}
\shortauthors{Gibson et al.}

\begin{document}

\title{An Automated Method for the Detection and Extraction of \hi\ 
	Self-Absorption in High-Resolution 21cm Line Surveys}

\author{Steven J. Gibson\altaffilmark{1}, 
	A. Russell Taylor\altaffilmark{1},
	Lloyd A. Higgs\altaffilmark{2},
	Christopher M. Brunt\altaffilmark{3},
	and Peter E. Dewdney\altaffilmark{2}}

\altaffiltext{1}{Dept. of Physics \& Astronomy, University of Calgary, 
2500 University Drive N.W., Calgary, Alberta T2N~1N4, Canada;
gibson@ras.ucalgary.ca; russ@ras.ucalgary.ca}

\altaffiltext{2}{Dominion Radio Astrophysical Observatory, 
Box 248, Penticton, British Columbia V2A~6K3, Canada;
Lloyd.Higgs@nrc.ca; Peter.Dewdney@nrc.ca}

\altaffiltext{3}{Department of Astronomy, LGRT 632, University of
Massachusetts, 710 North Pleasant Street, Amherst, MA 01003;
brunt@roobarb.astro.umass.edu
\vskip 0.35in
{\large \bf
\noindent For a version of this paper with full-resolution figures, please see 
\hspace*{0.35in} http://www.ras.ucalgary.ca/$\sim$gibson/hisa/cgps1\_survey/.
}
}

\begin{abstract}

We describe algorithms that detect 21cm line \hi\ self-absorption (HISA) in
large data sets and extract it for analysis.  Our search method identifies HISA
as spatially and spectrally confined dark \hi\ features that appear as negative
residuals after removing larger-scale emission components with a modified CLEAN
algorithm.  Adjacent HISA volume-pixels (voxels) are grouped into features in
$(\glon,\glat,\vel)$ space, and the \hi\ brightness of voxels outside the 3-D
feature boundaries is smoothly interpolated to estimate the absorption
amplitude and the unabsorbed \hi\ emission brightness.  The reliability and
completeness of our HISA detection scheme have been tested extensively with
model data.  We detect most features over a wide range of sizes, linewidths,
amplitudes, and background levels, with poor detection only where the
absorption brightness temperature amplitude is weak, the absorption scale
approaches that of the correlated noise, or the background level is too faint
for HISA to be distinguished reliably from emission gaps.  False detection
rates are very low in all parts of the parameter space except at sizes and
amplitudes approaching those of noise fluctuations.  Absorption measurement
biases introduced by the method are generally small and appear to arise from
cases of incomplete HISA detection.
This paper is the third in a series examining HISA at high angular resolution.
A companion paper (Paper~II) uses our HISA search and extraction method to
investigate the cold atomic gas distribution in the Canadian Galactic Plane
Survey.

\end{abstract}

\keywords{
	radiative transfer ---
	methods: analytical ---
	techniques: image processing ---
	surveys ---
	ISM: clouds ---
	ISM: structure 
}

\section{Introduction}
\label{Sec:intro}

The 21cm line of neutral atomic hydrogen (\hi) is a key probe of the Galactic
interstellar medium.  Although the cold ($T \la 100$~K) gas distribution is
difficult to map in \hi\ emission, \hi\ self-absorption (HISA) allows cold
foreground gas to be distinguished from warmer background gas at the same
radial velocity \citep{g02}.  Until recently, HISA has been studied in limited
low-resolution maps (e.g., \citealt{bb79,bl84}), or in a few isolated objects
at higher resolution (e.g., \citealt*{vdw88,f93}), but no detailed, systematic
surveys have been made.  High resolution allows feature structure to be
studied and unabsorbed background brightness to be estimated accurately.
Coverage of a wide area enables an unbiased look at the HISA population, e.g.,
without the a priori expectation that HISA is found only in molecular clouds.

High-resolution, wide-area HISA surveys have now become possible with the
advent of several major \hi\ synthesis surveys: the Canadian Galactic Plane
Survey (CGPS; \citealt{cgps}), the Southern Galactic Plane Survey (SGPS;
\citealt{sgps}), and the VLA Galactic Plane Survey (VGPS; \citealt{vgps}).
Past HISA studies have identified absorption features by eye, but this approach
is no longer adequate.  The very richness of the synthesis survey data sets
requires that they be analyzed in a rigorous, repeatable manner.  We have
therefore designed automated algorithms to identify and extract HISA features
from \hi\ longitude-latitude-velocity $(\glon,\glat,\vel)$ data cubes.

In this paper, we describe our HISA search and extraction algorithms.  We also
explain how we have tested our software with model data to determine its
reliability under a range of different conditions.  Large surveys are playing
an increasingly significant role in modern astrophysics, and it is essential
that their underlying methods are understood so their results can be
interpreted properly.  Following criteria established in \citeauthor{g00}
(\citeyear{g00}; hereafter Paper~I), our HISA search software seeks
finely-structured dark features against bright backgrounds that cannot be
confused with simple gaps in \hi\ emission.  Although its parameters are
optimized to identify HISA in the CGPS, the software is easily adapted to work
with other surveys (e.g., the VGPS: \citealt{g04}).

The CGPS uses a hexagonal grid of full-synthesis fields with single-dish
observations to enable the detection of all scales of \hi\ structure down to
the synthesized beam.  The CGPS \hi\ data have a $58\arcsec \times
58\arcsec{\rm cosec}(\dec)$ beam, 0.824~\kms\ velocity sampling, and a
field-center noise of $\trms \sim 3$~K in empty channels; $\trms$ doubles when
the 107\arcmin\ primary beam is filled with 100~K emission, and it can be up to
60\% greater between field centers.  The initial phase of the survey mapped a
$73\arcdeg \times 9\arcdeg$ region along the Galactic plane with longitudes
$74.2\arcdeg \le \glon \le 147.3\arcdeg$ and latitudes $-3.6\arcdeg \le \glat
\le +5.6\arcdeg$ ($+33.9\arcdeg \le \dec \le +68.4\arcdeg$), and extensions in
both $\glon$ and $\glat$ have followed.

Below, we describe our method of HISA identification and extraction at some
length (\S\ref{Sec:ext}) and evaluate the method's performance with models
(\S\ref{Sec:eval}).  A companion paper presents the results of our HISA search
of the $73\arcdeg \times 9\arcdeg$ Phase I CGPS (\citealt{g05a}; hereafter
Paper~II).  Subsequent papers in this series will apply the HISA search to
other data sets.

\section{Feature Extraction}
\label{Sec:ext}

\subsection{Identification Strategy}
\label{Sec:ext_strategy}

\subsubsection{Criteria}
\label{Sec:ext_strategy_crit}

Many HISA features are apparent to the eye (e.g., see
Figs.~\ref{Fig:real_hisa_id_maps_overview} -
\ref{Fig:real_hisa_id_stage_spectra}), but a complete visual search is unlikely
to be uniform, repeatable, or thorough, and it is also impractical given the
sheer volume and complexity of the CGPS data.  Thus, an automated search is
needed.  The search algorithm should find features meeting simple criteria that
can be confirmed by eye, but it should also be tested with model data to
evaluate its performance quantitatively (\S\ref{Sec:eval}).

The nature and appearance of HISA dictate how it can be identified.  First,
while the cold \hi\ from which it arises can have any extent, no HISA feature
can exceed the $(\glon,\glat,\vel)$ boundaries of its bright background \hi\
emission, or it ceases to be absorption.  Second, HISA must have different
$(\glon,\glat,\vel)$ structure than the background \hi\ for it to be
distinguishable from background fluctuations.  We choose to search for HISA
that is more finely-structured than the background \hi, since this is
consistent with the first constraint, most CGPS HISA that can be visually
identified is of this nature, and the exceptions (e.g., \citealt{kb01,k05})
are difficult to identify algorithmically.

We seek \hi\ features that can only be explained as HISA.  We prefer this
conservative approach over the alternative of including significant false
detections in our survey sample.  As given in Paper~I, our conditions for
distinguishing HISA from simple gaps in \hi\ emission are: (1) narrower line
widths than most observed emission features; (2) steeper line wings; (3) more
small-scale angular structure; and (4) a minimum \hi\ emission background
level.  The first two conditions are related for Gaussian line profiles, since
these have line wing slopes proportional to amplitude over width, but real HISA
need not be Gaussian.  The last condition excludes the finely-structured \hi\
emission gaps that are common at interarm velocities in the outer Galaxy, where
smooth, bright \hi\ backgrounds are often absent.  These four criteria exclude
HISA on larger angular and velocity scales or against weaker backgrounds, but
they are adequate for capturing most visible features.  We do not require the
extra condition of molecular line emission to confirm HISA features (e.g.,
\citealt{k74}), since many HISA features are visible without \twco\ emission in
the CGPS (Paper~I; \citealt{g02}) and in other surveys (e.g., \citealt{pb87}).

\subsubsection{Algorithms}
\label{Sec:ext_strategy_alg}

We tried and rejected many different methods before selecting the algorithm
described in this paper.  Discarded techniques include various derivative
measures to detect sharp edges, spatial and spectral curvature tests to look
for dips, Fourier and wavelet filtering methods, and unsharp masking.  Most of
these were successful in locating the strongest features, but few were robust
against noise, and many also produced large numbers of artifacts and false
detections.  The latter were especially frequent in methods that used only
spectral or spatial searches rather than both combined.

Our chosen method is based on a variant of the CLEAN algorithm \citep{h74}
developed by \citet*{sdi84} (hereafter the SDI CLEAN).  We remove large-scale
spectral and spatial emission structures from the \hi\ data iteratively and
flag the small-scale negative residuals as self-absorption features.  For
computational efficiency, these operations are carried out separately on the
spectrum at each spatial position $(\glon,\glat)$ and on the channel map at
each radial velocity $(\vel)$ in the data cube, and the results are combined
afterward.  The identified HISA is then subtracted from the \hi\ cube, and the
whole process is repeated until significant HISA can no longer be found; such
iteration allows features larger than the chosen CLEAN scales to be mapped.

The two spectral and spatial search algorithms are described below.  Each has
been tuned to find as much visually identifiable HISA as possible while
minimizing false detections.  The latter are further reduced by subsequently
requiring the HISA at any 3-D position $(\glon,\glat,\vel)$ to be detected by
both searches (\S\ref{Sec:ext_dt}).  The two algorithms were tuned by visually
comparing the search output against the observed \hi\ for many different
parameter value combinations, using a range of different HISA features with
different \hi\ emission backgrounds in the CGPS data.  The parameters that
yielded the most complete HISA detections with the fewest false detections were
used in the model-based search performance evaluations (\S\ref{Sec:eval}) and
in the CGPS HISA survey (Paper~II).

In the following discussions, the Galactic coordinate variables
$(\glon,\glat,\vel)$ are replaced by their pixel coordinate analogs $(i,j,k)$.
Adopted values for the search parameters are given in square brackets [~].
These give the best performance for a HISA search of CGPS data, but they may
not be universal.  In particular, the best filter scales and minimum background
level may differ for HISA searches elsewhere in the Galaxy.

\subsection{Spectral Search}
\label{Sec:ext_spectral}

At each spatial position $(i,j)$, the spectral search algorithm builds an
approximation of the ``unabsorbed spectrum'' $U(k)$ that would be observed if
no HISA were present.  
The algorithm assumes that $U(k)$ can be constructed from Gaussian functions of
a characteristic width $W_{char}$ that is narrower than the dominant emission
features but broader than the width of any expected HISA feature.  
Any channels in which the observed spectrum $O(k)$
deviates significantly negatively from $U(k)$ are flagged as possible HISA.

The iterative procedure used to derive $U(k)$ is a modification of the SDI
CLEAN.  $U(k)$ is initially set to zero, and the ``residual spectrum'' $R(k)$
is set equal to $S(k)$, a smoothed version of $O(k)$.  Smoothed data are used
to improve the signal-to-noise for the CLEANing process.  $S(k)$ is a {\it
spatial\/} average of $N \times N$ pixels centered at $(i,j)$, i.e., the
average of $O(i^\prime,j^\prime,k)$ intensities where $|i^\prime - i| \le
(N-1)/2$ and $|j^\prime - j| \le (N-1)/2$ [and $N = 7$ pixels $= 2.1$\arcmin].
Independent of this spatial averaging, the {\it spectral\/} \rms\ noise
$\sigma_{obs}$ in $O(k)$ is computed as the lowest of three \rms\ noise
measures over equal thirds of $O(k)$.  In the CLEAN loop, the following steps
are performed:

\begin{enumerate}

\item{If $R_{max}$, the peak value of $R(k)$, is less than a preset
fraction~[3\%] of the peak value of $S(k)$, the iteration ceases.}

\item{For any channel $k$ where $R(k)$ exceeds a given clip level~[0.8]
$\times R_{max}$, a ``correction spectrum'' $C(k)$ is set to a preset gain
[0.25] $\times R(k)$; elsewhere, $C(k)$ is set to zero.}

\item{$C(k)$ is convolved with a Gaussian whose full width at half maximum
(FWHM) is $W_{char}$ [8~\kms], and the resulting spectrum is added to $U(k)$.}

\item{The new residual spectrum is set to $R(k) = S(k) - U(k)$, and
$\sigma_{pos}$, the \rms\ of all {\it positive\/} values of $R(k)$ is computed
over all channels.  As $U(k)$ approaches $S(k)$, $\sigma_{pos}$ decreases; if
$\sigma_{pos} < \sigma_{obs}$, the iteration is terminated.}

\item{If the iteration has not terminated due to one of the above convergence
criteria, and a maximum number of loops [1000] has not been reached, steps 1-5
are repeated.}

\end{enumerate}

After the CLEAN loop is completed, adjacent channels where $R(k) < \sigma_{pos}
\, \times$ a factor $F$~[$-2.0$] are grouped into ``segments'' of suspected
HISA.  In each segment, $S(k)$ and $U(k)$ are evaluated at the channel
$k_{min}$ where $R(k)$ has a local minimum.  If $S(k_{min}) < 2 F
\sigma_{pos}$, the segment is rejected as a likely noise feature.  If
$U(k_{min}) < T_{crit}$, where $T_{crit}$ is a preset brightness level~[30~K],
the segment is rejected as having an insufficiently bright background to
identify HISA clearly.  A second, more conservative ${T_{crit}}^\prime$~[70~K]
is applied later when the spatial and spectral search results are combined
(\S\ref{Sec:ext_dt}).

A Gaussian profile is fitted to the absorption magnitude spectrum $R(k)$ of
each remaining channel segment.  Real HISA line profiles may not be Gaussian,
but this shape is assumed for simplicity.  For computational speed, the central
channel of the Gaussian is fixed at $k_{min}$, and the FWHM is fixed at one of
two values, $W_{narrow}$ [2.0 \kms] or $W_{broad}$ [4.0 \kms], to capture HISA
with a range of widths.  Model tests (\S\ref{Sec:eval}) show that many features
outside this range are also detected.  In each fit, a sloping linear base level
is derived along with the Gaussian amplitude $A$ and the standard deviation
$\sigma_{fit}$ of the fit.  A fit is rejected as statistically unreliable if
$A/\sigma_{fit} < D$~[2.0] or $A/\sigma_{pos} < D$.  It can also be rejected if
$O(k)$ lacks a morphological ``dip'' at $k_{min}$.  This is determined with a
filter function that returns a value of 1.0 for a dip between two equal peaks,
0.5 for a ``dip'' that drops only to the level of the adjacent spectral data on
one side of $k_{min}$, and 0.0 on a linearly rising or falling spectrum.  Fits
that return a value below the chosen threshhold [0.6] are rejected.  This
filter inhibits the detection of HISA on the edges of emission features that
are steeper than $W_{char}$ would allow; otherwise, significant false HISA
detections result.  For accepted fits, the channels in a {\it narrow\/} or {\it
broad\/} HISA spectrum (both initially zero) are set equal to the fitted
Gaussian if the amplitude exceeds a given fraction [5\%] of $A$.  Then, a
``detected'' HISA spectrum is created that consists of the maximum value in
each channel from the {\it narrow\/} and {\it broad\/} Gaussian fits to the
HISA line profile, to ensure full detection of the feature.  In the case where
the {\it broad\/} line wings do not correspond to real HISA in a narrow
feature, these will not be detected by the spatial search and will be removed
at a later stage of analysis.

Finally, when the HISA amplitudes have all been computed, these are spatially
smoothed [with a 1.5\arcmin\ beam] to join together groups of flagged
``flecks'' into more coherent features, and those that are sufficiently weak
and isolated are culled if their amplitude falls below a specified threshhold
[2~K].  An additional cosmetic improvement is made by excluding strong HICA
from the set of spectrally-identified HISA.  This is done by dropping any sight
lines from the search that contain channels whose continuum-subtracted line
brightness is significantly negative, i.e., if $O(k) < -6 \, \sigma_{obs}$.
Weaker HICA will survive this filter to contaminate the set of detected HISA.
Such contamination is difficult to remove in a way that leaves the ``pure''
HISA in the same sight lines intact.

\begin{figure}[ht]
\begin{minipage}[t]{16cm}
\centerline{\resizebox{15cm}{!}{\includegraphics[angle=0]{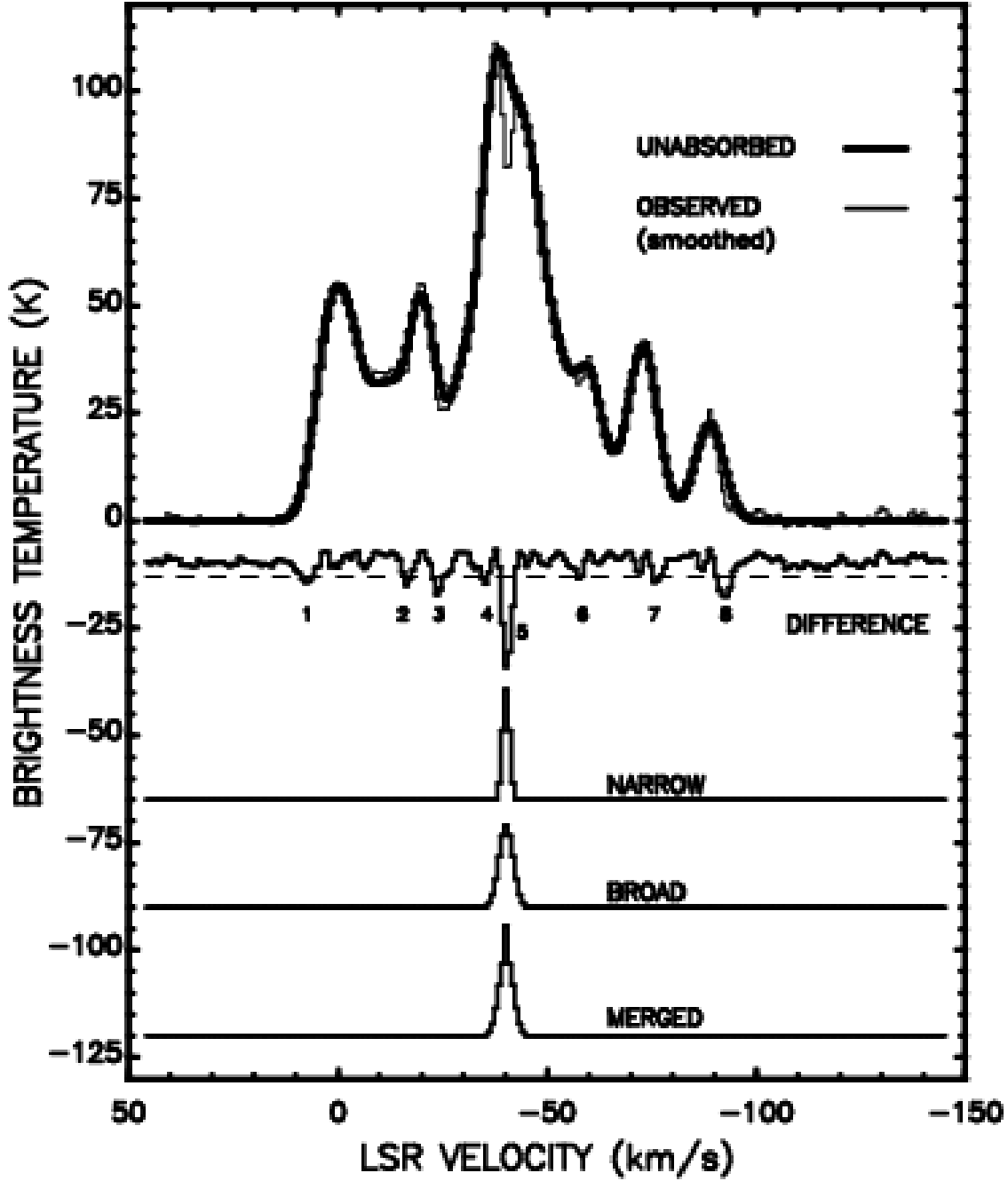}}}
\end{minipage}
\caption {
Velocity profiles showing HISA spectral detection stages for the Perseus HISA
Globule position.  The derived unabsorbed spectrum $U(k)$ and the spatially
smoothed observed spectrum $S(k)$ are shown in the top portion of the figure.
Below them is the residual or difference spectrum $R(k)$ (zero level at
$-10$~K).  The dashed line gives the level below which HISA is suspected.
Eight channel segments are indicated where this is the case.  Gaussian fitting
was accepted only for segment 5 (see text), and the resulting {\it narrow} and
{\it broad} HISA spectra are shown below this (zero levels at $-60$~K and
$-90$~K, respectively).  Finally, the ``detected'' HISA spectrum is shown at
bottom (zero level at $-120$~K).
}
\label{Fig:globule_spectrum_hi_dips}
\end{figure}

To illustrate the algorithm, we plot the \hi\ spectrum of Paper~I's Perseus
HISA Globule ($\glon = 139.635\arcdeg, \, \glat = 1.185\arcdeg$) in
Figure~\ref{Fig:globule_spectrum_raw} and several stages of the algorithm's
analysis in Figure~\ref{Fig:globule_spectrum_hi_dips}.  These \hi\ data
supersede those used in Paper~I, which contained a flaw that had no serious
impact on the results.  Appendix~\ref{App:erratum} gives further details.

In Figure~\ref{Fig:globule_spectrum_hi_dips}, the estimation of $U(k)$
converged when the peak residual became less than 3\% of the initial spectral
peak, giving $\sigma_{pos} = 3.6$~K (the dashed line indicates a negative
deviation of twice this value).  Suspected HISA was identified in eight channel
segments.  One of these (4) had no counterpart dip in $O(k)$.  Six others were
rejected because $U(k) < 70$~K.  For simplicity, we used $T_{crit} = 70$~K here
to show what would survive the ultimate ${T_{crit}}^\prime = 70$~K filter.  In
the remaining segment (5), channels in which the ``merged'' HISA spectrum
(maximum of the narrow and broad HISA spectra) is non-zero were then flagged as
having ``detected'' HISA.  Note that, because of the initial smoothing, the
detected HISA has a smaller amplitude than in
Figure~\ref{Fig:globule_spectrum_raw}.  The full HISA amplitude is recovered
when the spectral and spatial search results are merged (\S\ref{Sec:ext_dt}).

\subsection{Spatial Search}
\label{Sec:ext_spatial}

The spatial search algorithm is similar in principle to the spectral search,
although it does not attempt to fit absorption features with Gaussian shapes,
nor does it require that they satisfy a morphological ``dip'' filter.  It
begins by estimating the unabsorbed brightness distribution $U(i,j)$ in a given
spectral channel $k$.  The algorithm assumes that $U(i,j)$ can be constructed
from two-dimensional circular Gaussian components of a characteristic width
$G_{char}$ that is narrow enough to represent most \hi\ emission structure but
broader than any expected HISA features.  Clearly the choice of $G_{char}$
limits the angular size of HISA features that will be detected, although this
can be alleviated with repeated searches.

The iterative procedure used to derive $U(i,j)$ is again a modification of the
SDI CLEAN.  $U(i,j)$ is initially set to zero, and the residual map $R(i,j)$ is
set equal to $S(i,j)$, a spatially smoothed copy of the observed channel map
$O(i,j)$.  Use of $S(i,j)$, computed as an $N \times N$ pixel average of
$O(i,j)$ [with $N = 15$ pixels $= 4.5$\arcmin], improves the CLEAN convergence.
On a larger angular scale [20\arcmin], an estimate of the typical \rms\ noise
$\sigma_{obs}$ in $O(i,j)$ and its gross variation across the channel map are
derived in a manner similar to the spectral \rms\ noise in
\S\ref{Sec:ext_spectral}.  From this, the average \rms\ noise in $S(i,j)$,
termed $\sigma_{sm}$, and its variation over the map are deduced.  In the CLEAN
loop, the following steps are performed:

\begin{enumerate}

\item{$R_{max}$, the Mth [10th] highest value of $R(i,j)$, is found.  This is
chosen rather than the peak value so that the iteration process is not
dominated by one noisy pixel.  If $R_{max}$ is less than a preset fraction
[3\%] of the peak value of $S(i,j)$, the iteration ceases.}

\item{For any pixel $(i,j)$ where $R(i,j)$ exceeds a given clip level~[0.5]
$\times R_{max}$, a correction map $C(i,j)$ is set to a preset gain [0.25]
$\times R(i,j)$; elsewhere, $C(i,j)$ is set to zero.}

\item{$C(i,j)$ is convolved with a 2-D Gaussian whose FWHM is $G_{char}$
[20\arcmin], and the resulting image is added to $U(i,j)$.}

\item{The new residual map is set to $R(i,j) = S(i,j) - U(i,j)$, and the \rms\
value $\sigma_{pos}$ of the {\it positive\/} values of $R(i,j)$ is computed.
In deriving $\sigma_{pos}$, allowance is made for the fact that the noise may
vary across the image by applying suitable weights to the values of $R(i,j)$.
If $\sigma_{pos} < \sigma_{sm}$, the iteration is terminated.}

\item{If the iteration has not terminated due to one of the above convergence
criteria, and a preset maximum number of loops [1000] has not been reached,
steps 1-5 are repeated.}

\end{enumerate}

After the CLEAN loop is completed, all pixels where $R(i,j) < \sigma_{pos}(i,j)
\, \times$ a factor $F$~[$-2.0$] are noted, as are those where $O(i,j) - U(i,j)
< F \sigma_{obs}(i,j)$.  A map of ``suspected HISA'' is set equal to $U(i,j) -
O(i,j)$ for all pixels $(i,j)$ where either condition is met and zero
elsewhere.  This map is then filtered to remove pixels with amplitudes less
than a specified cutoff [4~K], as well as those for which $U(i,j) <
T_{crit}$~[30~K].
Lastly, as in the spectral algorithm, the suspected HISA map is smoothed [with
a 1.5\arcmin\ beam] to improve feature coherence, and a final cull is made of
smoothed amplitudes below a lower threshhold [2~K]; surviving features are
deemed ``detected HISA''.

\begin{figure}[ht]
\begin{minipage}[t]{16cm}
\centerline{\resizebox{15cm}{!}{\includegraphics[angle=0]{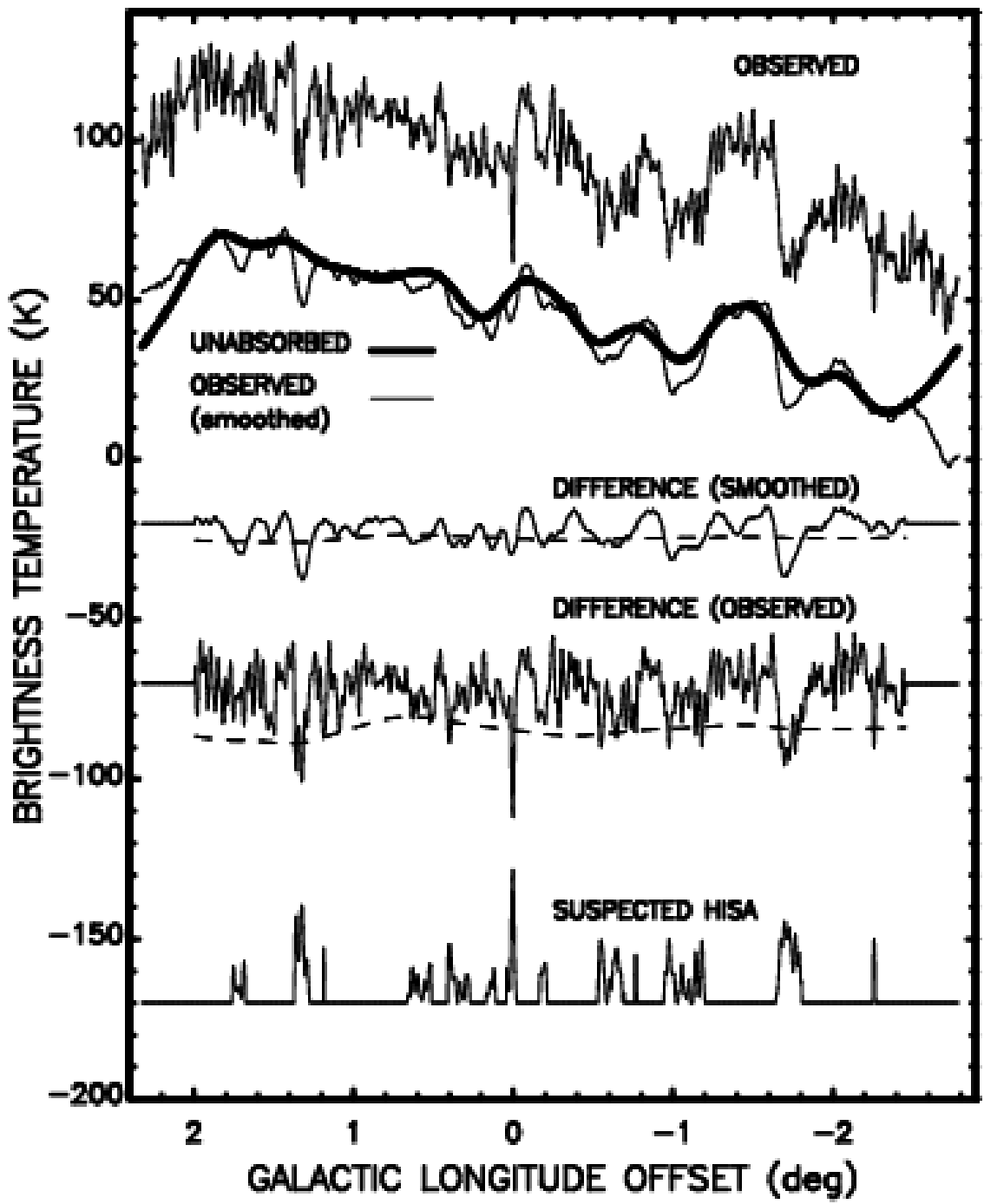}}}
\end{minipage}
\caption {
Latitude profiles showing HISA spatial detection stages for the Perseus HISA
Globule position (at which the Galactic latitude offset = 0\arcdeg).  From top
to bottom, cuts are taken through
the observed channel map $O(i,j)$, 
the smoothed observed map $S(i,j)$ and derived
unabsorbed map $U(i,j)$ (zero levels at $-50$~K),
the smooth difference map $S(i,j) - U(i,j)$ (solid)
and noise truncation level $-2 \sigma_{sm}$ (dashed) (zero level at $-20$~K),
the unsmooth difference map $O(i,j) - U(i,j)$ (solid)
and noise truncation level $-2 \sigma_{obs}$ (dashed) (zero level at $-70$~K),
and the suspected HISA features map prior to final amplitude culling
(zero level at $-170$~K).
Fast Fourier transforms (FFTs)
used in the CLEANing process make border areas of the 
map unusable after $S(i,j)$ and $U(i,j)$ are determined.
}
\label{Fig:globule_xcut_hi_abs}
\end{figure}

The spatial search is illustrated in Figure~\ref{Fig:globule_xcut_hi_abs}, with
longitude profiles taken through the Perseus HISA Globule position ($\glat =
1.185\arcdeg, \, \vlsr = -41.04~\kms$) from one channel map at different stages
of processing.  The determination of $U(i,j)$ took 114 iterations, ending when
$\sigma_{pos}$ became less than $\sigma_{sm} = 2.16$~K.

\subsection{HISA Amplitude Estimation}
\label{Sec:ext_dt}

\subsubsection{General Approach}

The physical properties of the absorbing gas cannot be understood without
knowing the HISA brightness temperature amplitude $\dt \equiv \ton - \tu$,
where $\ton(\glon,\glat,\vel) \equiv O(i,j,k)$ is the observed brightness on
the HISA feature, and $\tu(\glon,\glat,\vel) \equiv U(i,j,k)$ is the unabsorbed
emission that would be measured if no HISA were present.  Since only $\ton$ is
directly observed at the HISA position, $\tu$ must be estimated from $\toff$,
the \hi\ brightness off the HISA feature in space and/or velocity.  For
clarity, we note that $\tu$ represents {\it all\/} of the unabsorbed emission
along the line of sight at radial velocity $\vel$.  The emission from behind
the HISA feature that is subject to absorption is $\bfrac \cdot \tu$, where $0
< \bfrac \le 1$; the exact value of $\bfrac$ depends upon the sight-line
geometry (see Paper~I).

Several means of estimating $\tu$ have been used in past studies.  In the
spectral domain, the velocity edges of a HISA feature can be fitted with
straight lines \citep{hsf83,mbd95} or more complex functions
\citep{k74,msb78,lg03,k03} to estimate $\tu$ at intervening velocities.  In the
spatial domain, the \hi\ brightness at positions adjacent to the HISA feature
can be used directly as $\tu$ (Paper~I; \citealt{m01}) or as anchor points for
spatial fits across the feature \citep{f93,k03}.  A variant on this approach
assumes HISA is sufficiently diluted in the broad beam of a single dish
telescope to use the single dish spectrum at the feature position as $\tu$
\citep{vdw88}.

Our approach is more general.  We group the HISA volume-pixels (voxels)
identified in \S\S\ref{Sec:ext_spectral}-\ref{Sec:ext_spatial} into contiguous
3-D features in the spectral line cube.  For each feature, we estimate
$\tu$ by interpolating the $\toff$ values of the non-HISA voxels that
border the feature in $(\glon,\glat,\vel)$ space.  The interpolation uses a 3-D
Gaussian weighting function to ensure smoothness on the scale of the feature.
Specifically, at each position $(\glon,\glat,\vel)$ within the HISA feature,

\begin{equation}
\label{Eqn:off_interp}
\tu(\glon,\glat,\vel) \;\; = \;\; 
\frac{\displaystyle \sum_{n=1}^{\nedge}{\wnon \cdot \toff(\glon^{\prime}_n,\glat^{\prime}_n,\vel^{\prime}_n)}}
{\displaystyle \sum_{n=1}^{\nedge}{\wnon}}
\;\; ,
\end{equation}

\noindent where $n$ indexes the list of $\nedge$ off-HISA voxels with
coordinates $(\glon^{\prime}_n,\glat^{\prime}_n,\vel^{\prime}_n)$, and the
weight $\wnon$ is given by

\begin{equation}
\label{Eqn:off_weight}
\wnon \;\; = \;\; 
\exp \left[ 
- \frac{1}{2} \left( \frac{\glon^{\prime}_n - \glon} {\sigma_\glon} \right)^2
- \frac{1}{2} \left( \frac{\glat^{\prime}_n - \glat} {\sigma_\glat} \right)^2
- \frac{1}{2} \left( \frac{\vel^{\prime}_n  - \vel}  {\sigma_\vel}  \right)^2
\right]
\;\; .
\end{equation}

\noindent The Gaussian dispersions $(\sigma_\glon, \sigma_\glat, \sigma_\vel)$
are set so that each FWHM ($= \sigma \cdot \sqrt{8 \ln{2}}$) is half the
maximum length of any contiguous row of HISA voxels in that dimension, with
FWHM lower limits of 1.2\arcmin\ and 3.3\kms\ and upper limits of 20\arcmin\
and 8~\kms, the HISA search CLEAN scales.  The result is somewhat similar to
that of 1-D spectral fitting methods, but its structure is constrained by all
three dimensions of the \hi\ data.  This method yields $\tu$ and $\dt$
estimates superior to those of our separate spectral and spatial searches.

\subsubsection{Filtering}

The $\tu$ estimation algorithm considers two confidence levels of HISA.  First,
a {\it union\/} filter requiring a HISA identification from either the spectral
or spatial search is applied.  This filter includes nearly every HISA feature
that the eye can detect, as well as many non-HISA features that are discarded
later.  All accepted voxels are grouped into tentative HISA features and
interpolated over to obtain $\tu$, which is subtracted from the unsmoothed \hi\
data to get $\dt$ with the full CGPS angular resolution.  Then, an {\it
intersection\/} filter requiring HISA identification in {\it both\/} spectral
and spatial searches is applied.  Voxels not satisfying this filter are
unflagged as HISA, and their $\tu$ is reset to the observed \hi\ brightness.
Computing $\tu$ and $\dt$ for a union voxel set and applying an intersection
filter afterward ensures that (1) only the most likely HISA features survive,
and (2) any ``penumbral'' contamination from undetected HISA in their $\toff$
voxels is minimized; otherwise, $\tu$ and $|\dt|$ could be significantly
underestimated.  An alternative is to interpolate only over HISA satisfying the
intersection filter with all union-filter voxels dropped from the $\toff$
ensemble, but this frequently leaves too few edge voxels for a robust $\tu$
estimate.

Three additional filters are applied with the intersection filter.  Voxels with
$\dt \ge 0$ are discarded, as are those in the noisy peripheries of the survey
and those for which $\tu < {T_{crit}}^\prime$ [70~K], a stricter value than the
previous $T_{crit}$ [30~K] of \S\S\ref{Sec:ext_spectral}-\ref{Sec:ext_spatial}.
The peripheral culling rejects HISA voxels with CGPS field mosaic weights
$\wmos < 0.382$, the lowest weight that occurs between synthesis field centers.
Since $\wmos \propto {\sigma_{noise}^{-2}}$, this allows a maximum noise of
1.618 times the field center value (see \citealt{cgps}), which is typically
$5-7$~K for the $\tu \sim 70-130$~K levels of our HISA features.

The choice of ${T_{crit}}^\prime = 70$~K is empirically based.  As noted in
Paper~I and \S\ref{Sec:ext_strategy}, finely-structured \hi\ emission is common
in the CGPS data where the total amount of emission is low, i.e., off the plane
and at interarm velocities.  Without some sort of ${T_{crit}}$ filtering, the
HISA identification software is easily fooled in these regions, flagging many
false HISA features adjacent to and between sharp-edged emission.  We are
confident these are false HISA features, since the absorbing \hi\ would have to
be unrealistically cold to absorb against such faint \hi\ backgrounds, and such
apparently strong features are far less abundant in brighter emission fields
where they should be easier to detect and where more gas should be found
generally.  There is no single ${T_{crit}}^\prime$ value that excludes all such
false HISA while retaining all real HISA.  We chose ${T_{crit}}^\prime = 70$~K
to balance these two needs, with greater priority placed on the first.  For the
CGPS, 70~K rejects essentially all false HISA arising from sharp emission edges
while keeping most real HISA.  The model tests of \S\ref{Sec:eval} show that
the false HISA rejection is quite successful.  A few cases of some real HISA
being missed or truncated are discussed below and in Paper~II.

\subsubsection{Examples}

\begin{figure}
\begin{minipage}[t]{16cm}
\centerline{\resizebox{15cm}{!}{\includegraphics[angle=0]{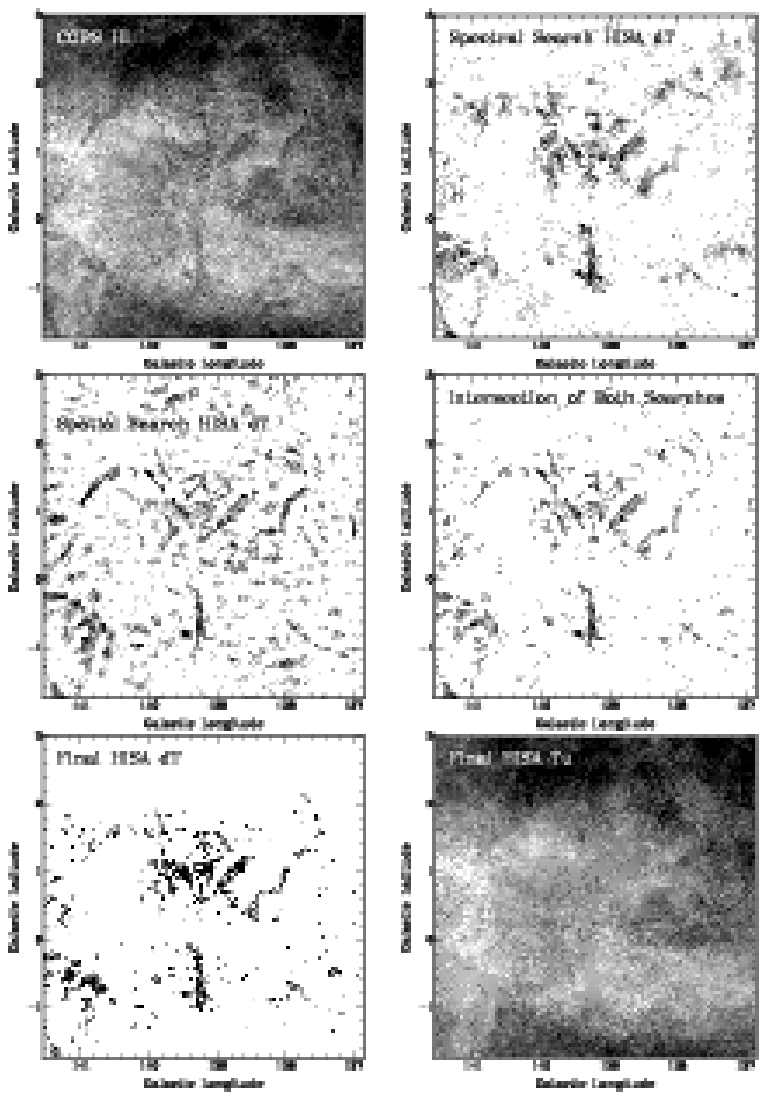}}}
\end{minipage}
\caption[]{
$(\glon,\glat)$ channel maps of sample Perseus HISA at $-41$~\kms, 
showing the same area as Figure~1 of Paper~I.  The panels give
CGPS \hi, HISA $\dt$ from the 
spectral search (\S\ref{Sec:ext_spectral}),
spatial search (\S\ref{Sec:ext_spatial}), and their intersection, 
and the final $\dt$ and $\tu$ from the full 3-D extraction
(\S\ref{Sec:ext_dt}).
Intensity ranges are $+40$ to 
$+130$~K for the first and last panels and $-20$ to 0~K 
for the rest, from black to white in all cases.
}
\label{Fig:real_hisa_id_maps_overview}
\end{figure}

Figures~\ref{Fig:real_hisa_id_maps_overview}-\ref{Fig:real_hisa_id_spectra}
illustrate the HISA amplitude extraction process with sample channel maps and
spectra that include the same features shown in
Figures~\ref{Fig:globule_spectrum_raw}-\ref{Fig:globule_xcut_hi_abs}.  These
give the initial \hi\ data, the $\dt$ amplitudes computed in different stages
of the analysis, and the final $\tu$.  As the figures show, most of the
visually apparent HISA is readily extracted.  Some residual HISA remains in
$\tu$, but its amplitude is a few K at most; the $\dt$ value of $-40$~K
extracted for the Perseus Globule is only 2~K weaker than that obtained with
the more conservatively-chosen Paper~I spatial $\toff$ boxes
(Table~\ref{Tab:cg_prop}).

\begin{figure}
\plotone{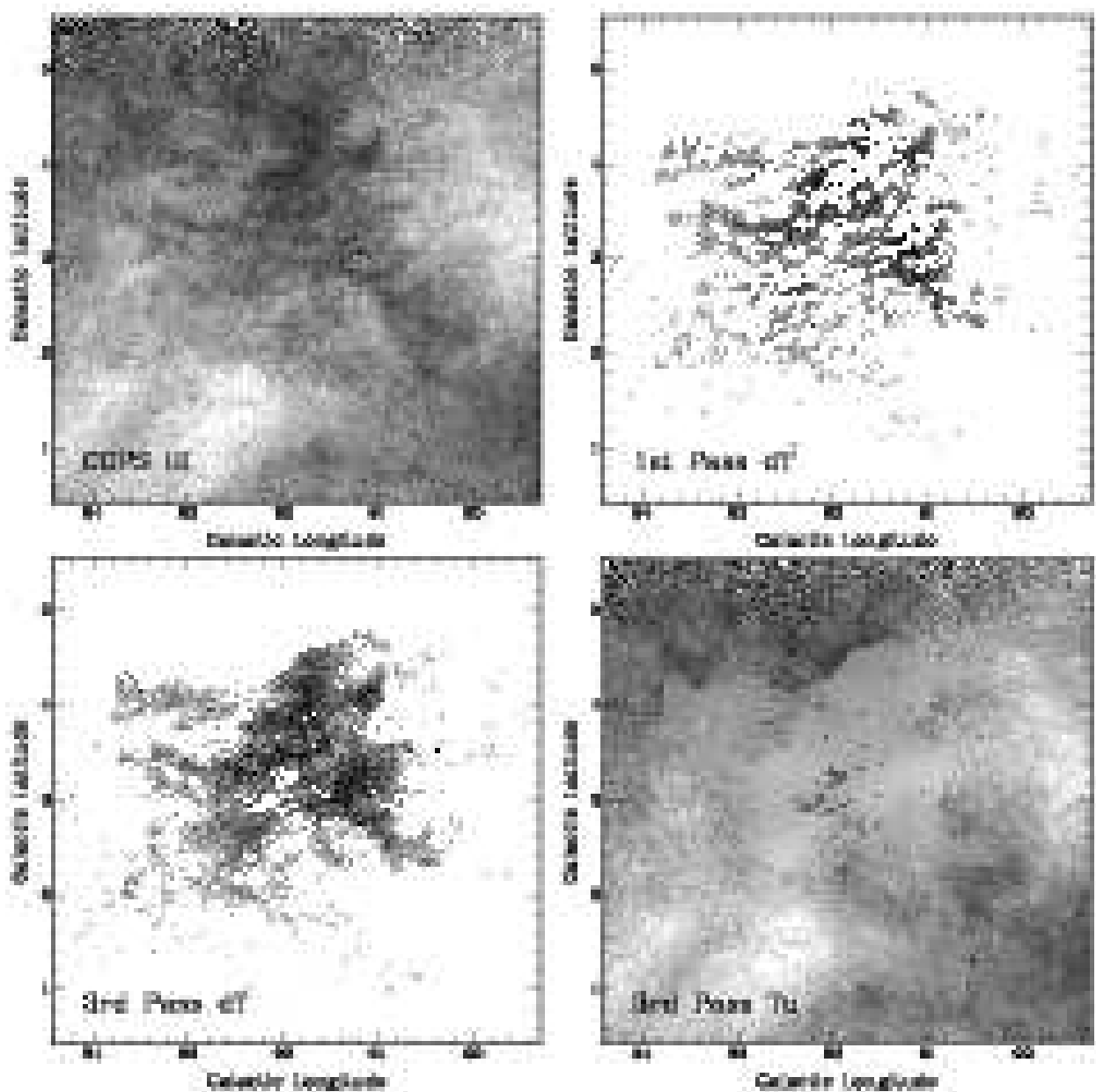}
\caption[]{
$(\glon,\glat)$ channel maps illustrating multiple-pass extraction of a large
HISA complex in the CGPS MK2 mosaic cube at at $-3$~\kms.  Shown are \hi\ 
brightness, first-pass $\dt$, third-pass $\dt$, and corresponding $\tu$.  
Intensity ranges are 0 to $+130$~K for the first and last panels and $-65$
to 0~K 
for the two $\dt$ maps.
The small $6\arcmin \times 6\arcmin$ box ($\glon=91.20\arcdeg, 
\glat=+2.97\arcdeg$) marks the area from which the  
spectra in Figure~\ref{Fig:real_hisa_id_stage_spectra} were extracted.
The feature extraction is truncated for $\glat \ga +4.5$\arcdeg\ due to 
$\tu < 70$~K (see \S\ref{Sec:ext_dt}).
}
\label{Fig:real_hisa_id_stage_maps}
\end{figure}

\begin{figure}
\plotone{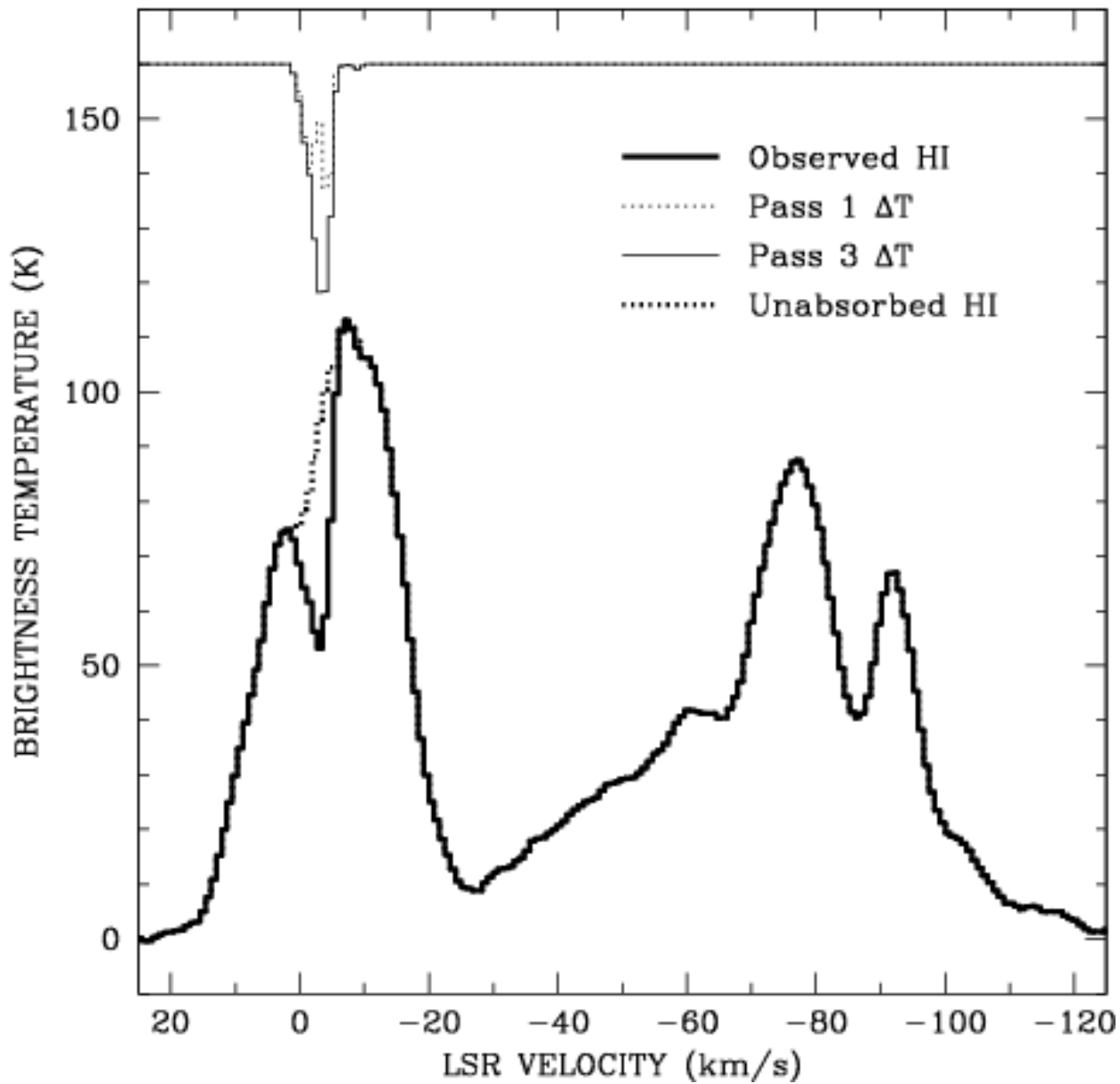}
\caption[]{
Spatially-averaged velocity spectra illustrating multiple-pass HISA extraction.
The spectra are extracted from the $6\arcmin \times 6\arcmin$ box marked in 
Figure~\ref{Fig:real_hisa_id_stage_maps} at 
$(\glon=91.20\arcdeg, \glat=+2.97\arcdeg)$.
Shown are \hi\ emission, first-pass HISA $\dt$, third-pass HISA $\dt$, and 
third-pass HISA $\tu$.
For clarity, the $\dt$ zero-points have been shifted to 160~K.
}
\label{Fig:real_hisa_id_stage_spectra}
\end{figure}

The HISA identification software cannot detect features larger than its CLEAN
scales [8~\kms\ and 20\arcmin] and instead flags only their darkest parts.  To
overcome this limitation, we feed the $\tu$ cubes back into our search
algorithms to identify HISA missed on the previous pass
(\S\ref{Sec:ext_strategy}).  Subsequent $\dt$ extractions are made with unions
of HISA flags from all prior search passes but always use the original \hi\
data for $\toff$.  Three such passes are adequate for the CGPS \hi\ data set.
Of the HISA voxels extracted in all three passes, 86.5\% were found in the
first pass, 11.2\% in the second, and only 2.3\% in the third.
For brevity,
Figures~\ref{Fig:real_hisa_id_maps_overview}-\ref{Fig:real_hisa_id_spectra}
display only third-pass results, which differ little from the first pass for
this example.  A case with more dramatic differences between passes is
illustrated in Figures~\ref{Fig:real_hisa_id_stage_maps}~\&
\ref{Fig:real_hisa_id_stage_spectra}.  Features this big require multiple
passes to capture.  HISA flagging is significantly improved after the first
pass in both the spatial and spectral domains.  We show only first- and
third-pass results here, since the second pass closely resembles the third.
After three passes, HISA flagging is incomplete in only a few places due to
$\tu < 70$~K truncation, mostly near the northern edge of the map
(Fig.~\ref{Fig:real_hisa_id_stage_maps}).  
Aside from minor losses from $\tu$ changes, the flagged HISA generally
increases, with the fraction of the total flagged per pass being 72.9\%,
22.5\%, and 4.6\% for passes 1, 2, and 3.  The smooth $\tu(\glon,\glat,\vel)$
structure in Figures~\ref{Fig:real_hisa_id_stage_maps} \&
\ref{Fig:real_hisa_id_stage_spectra} shows that our $\tu$ estimation method
follows the large-scale \hi\ emission brightness reasonably well.

\subsubsection{A Note on the Assignment of Structure}

We have chosen to attribute fine-scale structure in $\ton$ to $\dt$, leaving
$\tu$ smooth on the scale of the HISA feature.  This approach presumes that the
absorbing gas is finely structured and the \hi\ background is not, consistent
with our adopted HISA identification strategy (\S\ref{Sec:ext_strategy}).  Such
consistency allows the identified HISA structure to be removed so that
subsequent search passes do not flag it again.  If however some $\ton$
structure arises from $\tu$ (e.g., \citealt{kb01}), the true $\dt$ is smoother
than we have found.  We feel that our choice of method is reasonable for most
circumstances.  Small-scale \hi\ emission structure is common in the general
ISM but appears minimized in the bright, smooth \hi\ fields where we see most
CGPS HISA.

\section{Survey Reliability and Completeness}
\label{Sec:eval}

\subsection{Motivation}
\label{Sec:eval_mot}

The eye is the first means of identifying HISA features that meet the
appropriate criteria, and the search algorithms of \S\ref{Sec:ext} were
designed primarily to mimic visual detection.  However, the eye can be fooled;
for example, it often finds false patterns in noise, perhaps due to
evolutionary pressures to spot predators \citep{p93}.  We made our HISA search
and extraction algorithms as rigorous as possible, but they remain limited by a
number of factors, including:

\begin{enumerate}

\item {\bf $\tu$ faintness:} To avoid confusion with emission gaps at low
column densities, HISA detection is blocked if $\tu < 70$~K.  Where this
occurs, small features or parts of large ones may be missed.

\item {\bf $\tu$ underestimation:} We assume $\tu$ is not finely structured and
estimate it from $\toff$ voxels surrounding the HISA feature in 3-D.  However,
many HISA features occur near spectral emission peaks.  If $\tu > \toffavg$,
then we underestimate $\tu$ and $|\dt|$.  Both can also be underestimated if
HISA flagging is incomplete and an unidentified ``penumbra'' of faint HISA
contaminates $\toff$.

\item {\bf Noise degradation:} Despite smoothing, some low-amplitude HISA will
be lost to noise.  Whole features may be missed, or just their cores may be
detected, making them appear smaller, clumpier, and more fragmented than they
really are.  In addition, false HISA detections will be introduced by noise
fluctuations at low $|\dt|$.

\item {\bf Overlarge Structure:} By design, the HISA search algorithms cannot
flag whole features larger than the adopted $20\arcmin$ and $8\kms$ CLEAN
filter scales.  The use of multiple search passes eases this limitation but may
not remove it entirely.

\item {\bf Unresolved Structure:} Small-scale HISA structure may be diluted or
missed entirely if undersampled.  Angular structure down to the $1\arcmin$ CGPS
beam is seen (Paper~I), so smaller-scale structure seems likely.  HISA
linewidths narrower than the CGPS Nyquist limit of $1.65\kms$ also exist
\citep{k74,lg03}.  Such linewidths are rare in random HICA sight lines (e.g.,
\citealt*{cst88}), but since continuum backgrounds can be brighter than \hi\
backgrounds, HICA can include warmer absorbing gas than HISA, and it's possible
that HISA lines may be narrower on average.  HISA velocity dilution is thus a
real concern in synthesis surveys, while angular dilution will be more severe
for single-dish telescopes; no present instrumentation can adequately sample
both the angular and velocity structure of HISA.

\end{enumerate}

To evaluate such limitations objectively and quantitatively, we tested our
software's ability to extract HISA features from model \hi\ data.  Our goal was
to understand (1) what fraction of HISA the software detects, (2) how many
detections are false positives, and (3) how much the detected features differ
in size and amplitude from their input versions.

\subsection{Models}
\label{Sec:eval_mod}

The model 21cm spectral line cubes were sums of noisy, positive-amplitude
emission backgrounds and noise-free, negative-amplitude absorption features.
Gas properties and radiative transfer effects were not considered, as these are
irrelevant to the detection software's performance.  To test this under varying
conditions, 64 randomly-configured model cubes were made.  Each cube used the
standard CGPS pixel and channel sizes, with dimensions half those of a standard
CGPS mosaic cube for computational efficiency: $2.56\arcdeg \times 2.56\arcdeg
\times 106\kms$ ($512 \times 512 \times 128$ voxels).  Sample model data are
shown in Figure~\ref{Fig:hockey_pucks}.

\subsubsection{Absorption Features}
\label{Sec:eval_mod_abs}

Each model HISA feature was given a cylindrical shape in the \hi\ line cube,
with a Gaussian velocity profile and a flat-disk spatial profile convolved with
a 60\arcsec\ circular beam.  Although simple, these angular and velocity
profiles are similar enough to typical HISA for testing purposes.  The
features, known as ``hockey pucks'' for their usually oblate aspects in the
CGPS voxel grid, are parameterized by their unconvolved angular FWHM $\puckda$,
velocity FWHM $\puckdv$, and (negative) central amplitude $\puckdt$.

2048 hockey pucks were inserted into each model cube with random sizes,
amplitudes, and positions.  The $(\glon,\glat,\vel)$ and $\puckdt$
distributions were uniformly random, except that puck overlaps in
$(\glon,\glat,\vel)$ were prevented, with a minimum separation of 1 voxel
enforced between pucks at an absorption threshhold of 0.005~K.  The $\puckda$
and $\puckdv$ distributions were skewed toward small features, with relative
probabilities of $P(\puckda) \propto \puckda^{-2}$ and $P(\puckdv) \propto
\puckdv^{-1}$.  This was done to counter the fact that larger pucks have more
voxels.  We measured angular and velocity widths locally from each voxel in the
performance analysis (\S\ref{Sec:eval_anal}), and $P(\puckda)$ and $P(\puckdv)$
made our voxel-based size distributions more evenly sampled.
The puck parameter ranges used were $0.1\arcmin \le \puckda \le 60\arcmin$,
$0.355\kms \le \puckdv \le 16.0\kms$, and $-1$~K $\ge \puckdt \ge$ $-40$~K.
$\puckda=0.1\arcmin$ results in ``unresolved'' structures that get diluted in
the CGPS beam
Similarly, $\puckda = 0.355$~\kms, which would occur for purely thermal \hi\
linewidths at 2.73~K, would be unresolved by the CGPS in velocity.  Both cases
test the detection limits for fine-scale HISA structure.  At the other extreme,
the software's sensitivity to structures larger than the 20\arcmin\ and 8\kms\
CLEAN filter scales is also tested.

\subsubsection{Emission Background}
\label{Sec:eval_mod_emis}

The background emission fields were similarly constructed of random ensembles
of cylindrically-symmetric components.  These differed from the hockey puck
absorption features in that they had positive amplitudes, simple Gaussian
angular profiles, and minimum sizes equal to the CLEAN scales.  They were also
allowed to overlap and fill the entire cube, so we refer to them as emission
components rather than discrete features.  Size ranges were $20\arcmin \le
\epuckda \le 120\arcmin$ and $8\kms \le \epuckdv \le 20\kms$, with
$P(\epuckda)$ and $P(\epuckdv)$ the same as for the HISA pucks.  The amplitude
range was $+1$~K $\le \epuckdt \le$ $+20
\left[\epuckvol/\epuckvolmax\right]$~K, where the component volume $\epuckvol
\equiv \epuckda^2 \epuckdv$, $\epuckvolmax = (120\arcmin)^2 \, 20\kms$, and the
$\epuckdt$ distribution was further skewed as $P(\epuckdt) \propto
\epuckdt^{-1}$.  These adjustments placed most of the power at large scales, as
is seen in real \hi\ emission \citep{g93}.  4096 components were summed to make
each model cube's emission field.  This was subsequently rescaled to give a
median brightness temperature of 70~K, so that half the cube on average would
allow HISA detections, and $\tu \sim 70$~K effects could be easily studied.

For greater realism, noise was added to the \hi\ model.  A 3-D field of
uncorrelated Gaussian random voxel noise was convolved with a 60\arcsec\ FWHM
Gaussian beam and a 1.319\kms\ FWHM Gaussian velocity point spread function
(PSF) to mimic the structure of correlated noise in the CGPS data, and the
\rms\ noise amplitude was scaled to match the 6~K level found in CGPS field
centers filled with 100~K emission.  Unlike the CGPS noise, the model noise
does not vary with distance from field centers, its beam is
declination-independent, and its velocity PSF is not the true CGPS velocity
PSF, which is the Fourier transform of a Gaussian truncated at 20\% of peak
amplitude, with an effective FWHM of $1.319~\kms = 1.6$ channels.  However,
none of these differences should seriously affect the performance analysis.

\subsection{Analysis}
\label{Sec:eval_anal}

The same procedures used to search for HISA in the CGPS data (\S\ref{Sec:ext})
were applied to the model data.  The software performance was then evaluated by
comparing the input and extracted absorption.  Each of the 64 model cubes was
analyzed separately, and the results were merged afterward to maximize coverage
of the model parameter space.

\subsubsection{Measurement of HISA Observables}
\label{Sec:eval_anal_param}

Four observables were extracted from the HISA data: the absorption amplitude
$\dt$, unabsorbed brightness $\tu$, angular width $\da$, and velocity width
$\dv$ of the absorption.  All four were measured at each voxel
$(\glon,\glat,\vel)$ position rather than on a per-feature basis, because the
CGPS HISA has complex structure, and properties can vary within one feature.
However, $\da$ and $\dv$ are still aggregate properties that depend on the
local distribution of HISA around them.

\begin{figure}
\plotone{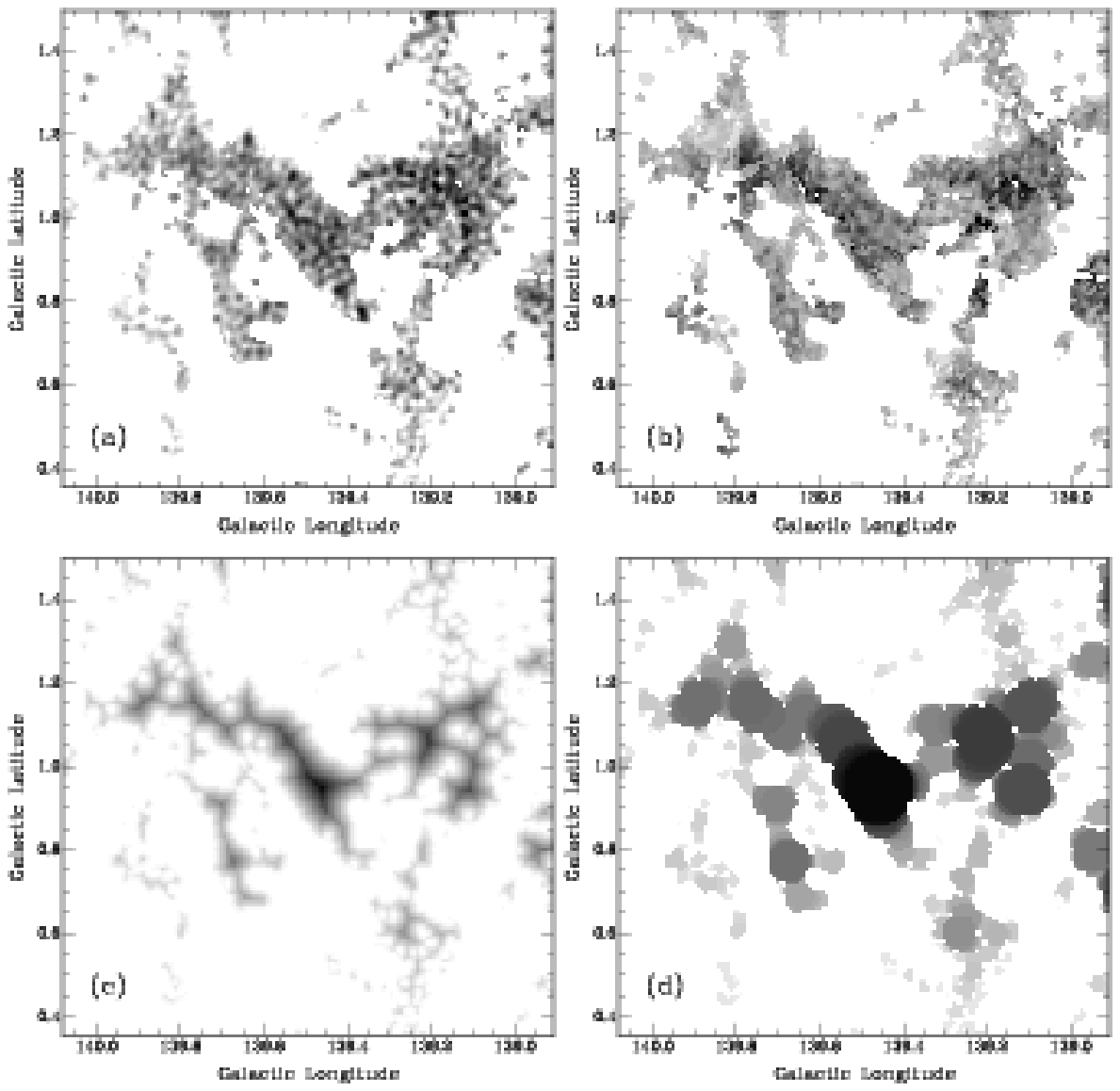}
\caption[]{
Channel maps illustrating velocity width and angular width measures:
{\it (a)\/} HISA absolute amplitude $|\dt|$; 
{\it (b)\/} $\dv$, the line full width at half maximum;
{\it (c)\/} $2\times \aoff$, where $\aoff$ is the 
offset to the nearest HISA feature edge;
and
{\it (d)\/} $\da$, the angular width obtained from $2 \, \aoff$ values imposed out to a radius $\aoff$.
Intensity ranges are linear, from white to black, for 
$0$~K $\le |\dt| \le 40$~K,
$0\kms \le \dv \le 6\kms$,
$0\arcmin \le 2 \aoff \le 10\arcmin$, and
$0\arcmin \le \da \le 10\arcmin$.
See \S\ref{Sec:eval_anal_meas} for further details.  
}
\label{Fig:da_dv_examples}
\end{figure}

The velocity width $\dv$ measures the line FWHM.  For each HISA voxel, all HISA
contiguous in $\vel$ at the same $(\glon,\glat)$ position is examined to find
the channel with maximum $|\dt|$.  On either side of this channel, the closest
channels for which $|\dt| \le 0.5 \, |\dt|_{max}$ are identified; non-HISA
channels with $\dt = 0$ are included if necessary.  The half-maximum velocities
are refined to sub-channel accuracy by linear interpolation.  The difference
between them is $\dv$.  This $\dv$ is assigned to all HISA voxels in the same
velocity grouping at the same $(\glon,\glat)$.  We make no attempt to correct
for instrumental broadening (e.g., \citealt{mbd95}), since this is nontrivial
for the complex line structure of some HISA, and only the narrowest features
will be broadened significantly in the CGPS.  Figure~\ref{Fig:da_dv_examples}
shows a sample map of $\dv$.  On average, the broader linewidths occur in
larger HISA features.

The angular width $\da$ measures the diameter of the largest circle containing
the $(\glon,\glat)$ position and zero non-HISA voxels at the same velocity.
This scheme measures the edge-to-edge feature width on a local scale.  Unlike
the FWHM-based $\dv$, $\da$ uses the full HISA extent.  Experiments with an
angular FWHM proved too sensitive to complex internal structure in the
$\dt(\glon,\glat)$ distribution to be interpreted easily.  The resulting $\da$
measures can be a bit larger than the FWHM-based hockey puck $\puckda$,
especially for large puck amplitudes $\puckdt$, but since the same $\da$
measure is taken of the HISA model inputs and outputs, the method is internally
consistent.

Figure~\ref{Fig:da_dv_examples} illustrates how $\da$ is measured.  From each
HISA voxel, the angular offset $\aoff$ to the nearest non-HISA voxel with the
same velocity is found.  $\aoff(\glon,\glat)$ maps HISA ``skeletons'' whose
ridge-like maxima equal half the local width of the feature.  To build a $\da$
map, we step over all $(\glon,\glat)$ positions and write $2 \,
\aoff(\glon,\glat)$ to all points $(\glon^\prime,\glat^\prime)$ in a new map
for which $\sqrt{\left(\glon^\prime - \glon\right)^2 + \left(\glat^\prime -
\glat\right)^2} \le \aoff(\glon,\glat)$, where the largest imposed $2 \, \aoff$
value is always retained.  This yields the angular width of HISA filamentary
structure at that velocity.

\subsubsection{Performance Measures}
\label{Sec:eval_anal_meas}

The HISA extraction software's performance was measured in three ways: the
``throughput'' fraction $\fid$ of model HISA detected; the ``true fraction''
$\fot$ of HISA detections corresponding with model input; and the ``drifts''
$\ddt$, $\dtu$, $\dda$, and $\ddv$ between input and output properties.  All
were measured as functions of the HISA observables $(\dt,\tu,\da,\dv)$, which
define a 4-D parameter space in which the software performance is evaluated.

The $(\glon,\glat,\vel)$ positions and $(\dt,\tu,\da,\dv)$ properties were
first tabulated for all HISA voxels in both the input and output model
$(\glon,\glat,\vel)$ cubes.  The 4-D voxel count histograms
$\via(\dti,\tui,\dai,\dvi)$ and $\voa(\dto,\tuo,\dao,\dvo)$ were constructed
from these voxel tables, using bin dimensions of 2.5~K $\times$ 2.5~K $\times$
0.5\arcmin\ $\times$ 0.5\kms.  
In parallel, the voxel count histogram $\vid$ was made from all input voxels
with output at the same $(\glon,\glat,\vel)$, and $\vot$ was made from all
output voxels with input at the same $(\glon,\glat,\vel)$.  Throughputs and
true fractions were then derived as $\fid \equiv \vid / \via$ and $\fot \equiv
\vot / \voa$.  From the subset of voxels appearing in both the input and output
HISA cubes, four 4-D drift histograms of the average changes undergone by
$\dti$, $\tui$, $\dai$, and $\dvi$ as functions of $(\dti,\tui,\dai,\dvi)$ were
assembled, e.g., as $\ddt \equiv \langle \dto - \dti \rangle$, with the average
taken over all HISA voxels in $(\glon,\glat,\vel)$ with the same
$(\dti,\tui,\dai,\dvi)$ properties.

The 4-D performance histograms were computed for all 64 HISA models, which were
identical apart from different random number inputs.  The results were merged
together into a single set of histograms and smoothed with 4-D Gaussians to
improve the performance measure reliability and coverage of the parameter
space.  Variable smoothing scales were used, because the parameter space
coverage was sparser in some areas than in others.  The smoothing FWHM were
$0.5 |\dt|$, $0.5 |\tu - 70~{\rm K}|$, $0.5 \da$, and $0.5 \dv$, with minimum
values of 6.0~K, 6.0~K, 1.0\arcmin, and 1.319\kms\ to match the model $T_{rms}$
and CGPS resolution.  This scheme preserved structure in the well-sampled parts
of the parameter space and interpolated it smoothly elsewhere.

$\voa(\dto,\tuo,\dao,\dvo)$ histograms were also computed for all 36 mosaic
cubes of real CGPS HISA and summed together to assess the distribution of
observed HISA in the survey.  As with the HISA feature catalog in Paper~II,
sight lines with $\tc > 20$~K were excluded.

\subsection{Results}
\label{Sec:eval_res}

\subsubsection{Model Parameter Distributions}
\label{Sec:eval_res_dist}

\begin{figure}
\begin{minipage}[t]{16cm}
\centerline{\resizebox{15cm}{!}{\includegraphics[angle=0]{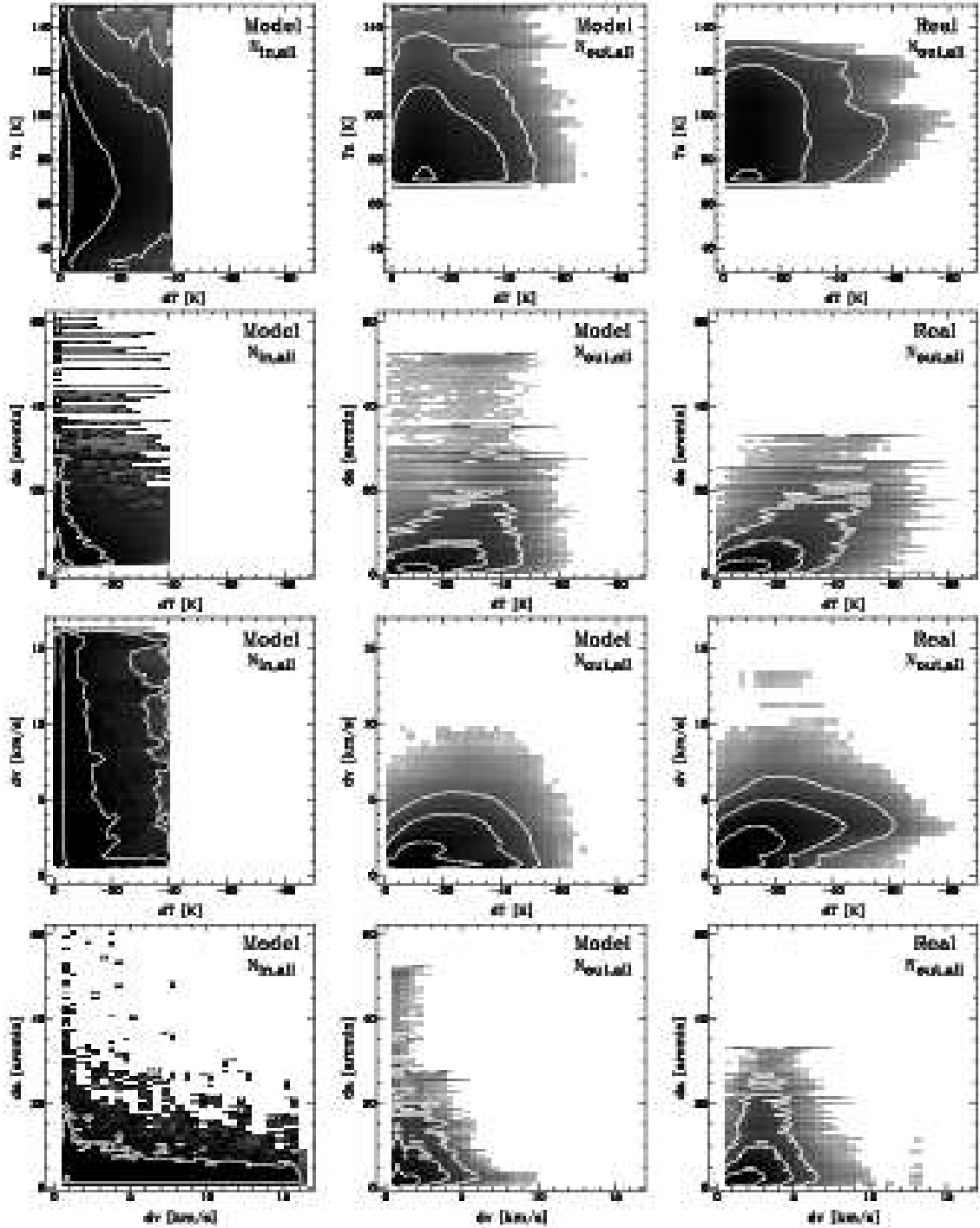}}}
\end{minipage}
\caption[]
{
2-D projections of 4-D property histograms of input voxels (model $\via$),
extracted voxels (model $\voa$), and observed CGPS HISA (real $\voa$).  Axis
labels are ``dT'' = $\dt$, ``Tu'' = $\tu$, ``da'' = $\da$, and ``dv'' = $\dv$.
Counts were summed along the orthogonal axes, so the full distributions are
visible.  No significant trends in $(\tu,\da)$ or $(\tu,\dv)$ were found.  The
intensity scale is logarithmic from 1 count (light) to 1 million counts (dark).
Contours mark counts of $10^3$, $10^4$, $10^5$, $10^6$, and $10^7$ in all
panels except the $(\dt,\da)$ and $(\dv,\da)$ maps of the model $\via$, where
$10^3$ and $10^4$ are omitted.
}
\label{Fig:4hist_counts}
\end{figure}

To examine general $(\dt,\tu,\da,\dv)$ parameter distributions and trends, we
made 2-D projections of the unsmoothed model $\via$ and $\voa$ and the real
CGPS $\voa$ by summing the counts along the 2 other axes in the 4-D parameter
space.  A number of these 2-D projections are shown in
Figure~\ref{Fig:4hist_counts}.  

The input models fill ranges of $0 > \dt > -40$~K, $30 < \tu < 150$~K, $0 < \da
< 61\arcmin$, and $0.8 < \dv < 16$\kms, with peaks at $\tu = 70$~K and small
$|\dt|$ and $\da$; the $\dv$ distribution was relatively flat.  The peaked
$\dt$ and $\da$ distributions occurred despite the shapes chosen for the puck
property distributions (\S\ref{Sec:eval_mod}).  The low-$|\dt|$ peak is due to
faint HISA in feature line wings and spatial envelopes.  The low-$\da$ peak
results from $\da$ being measured from HISA voxels above a minimum $|\dt|$,
which makes pucks appear smaller off the line center.  In a similar way, pucks
with the same $\puckda$ have greater $\da$ if $|\puckdt|$ is larger, and no
voxels with $\da < 1.5$\arcmin\ and $\dt < -20$~K are found.

The extracted model $\dt$ peak is shifted to $\sim -10$~K.
A tail of strong absorption extends to $\dt \sim
-60$~K.  Although they account for only 2\% of the total HISA voxels, these
$\dt < -40$~K points demonstrate that some $\dt$ drift occurs.  The extracted
$\tu$ is truncated at $\la 70$~K but otherwise appears unchanged from the input
model.  Large $\da$ and $\dv$ values are both truncated as predicted in
\S\ref{Sec:eval_mot}, although less severely for $\da$, since the non-Gaussian
angular profiles better survive the CLEAN process.  Iterative extraction
(\S\ref{Sec:ext}) allows much of the puck structure $> 20$\arcmin\ to be
recovered here, but the purely Gaussian puck velocity profiles with $\puckdv >
8$\kms\ are CLEANed out of the data with great efficiency.  $\da$ peaks at the
same location as the input data but is more concentrated.  $\dv$ is also now
concentrated toward low values.

\subsubsection{Real Parameter Distributions}
\label{Sec:eval_res_dist_real}

The CGPS HISA $\dt$ has a larger range than in the extracted models, due to a
few very strong features like GHISA 079.88$+$0.62$+$02 and GHISA
091.90$+$3.27$-$03 (see Paper~II for feature details).  Its $\dt$ peak is
similar to the models'.  The CGPS HISA is truncated for $\tu < 70$~K as well as
$\tu > 135$~K, where the maximum \hi\ brightness is reached.  $\da$ peaks at
the same scale as the model output but has a lower maximum scale, perhaps
because real HISA is more porous.  The model output $\dv$ range is slightly
exceeded.  The $\dv$ peak and maximum value are both a little broader than for
the extracted model HISA.

The CGPS HISA fills almost the same parameter space as the extracted model
HISA.  Since some model parameter ranges are larger on input than output, the
ranges of real HISA properties may exceed those observed in the CGPS.  HISA
with $\tu < 70$~K is already known (e.g., \citealt{kb01}), and HISA with $\dv >
8$\kms, $\da > 33$\arcmin, or $|\dt| > 80$~K is possible, although $|\dt| \le
\bfrac \, \tu + \tc - \ts$ is required by the 4-component radiative transfer
equation (Paper~II, Eqn.~1), where $\ts$ is the spin (excitation) temperature
of the absorbing gas.  Also, $\da$ is limited by feature porosity.  GHISA
091.90$+$3.27$-$03 exceeds the $\da$ limit in gross extent but is not
completely solid.  Larger features are known (e.g., the \citealt{rc72} ``cold
cloud'' toward the Galactic center), but their porosity at 1\arcmin\ resolution
has not been reported.

The input models had no built-in correlations of feature properties, and the
same is largely true for the real HISA.  Certainly $\dt$ and $\tu$ are not
related in the CGPS, except that the strongest $\dt$'s prefer some $\tu$ values
over others.  However, the peak CGPS $\da$ and $\dv$ both increase gradually
with $|\dt|$ out to $\dt = -40$~K, $\da = 20$\arcmin, and $\dv = 4$\kms.  These
trends have considerable scatter, and there are weaker versions in the
extracted model HISA.  But if they reflect real HISA behavior, then stronger
absorption is more likely to have larger contiguous angular structure or
broader linewidths, although $\da$ and $\dv$ do not correlate as well with each
other as they do with $\dt$.

\subsubsection{Throughput and True Fraction}
\label{Sec:eval_res_fid_fot}

\begin{figure}
\begin{minipage}[t]{16cm}
\centerline{\resizebox{10cm}{!}{\includegraphics[angle=0]{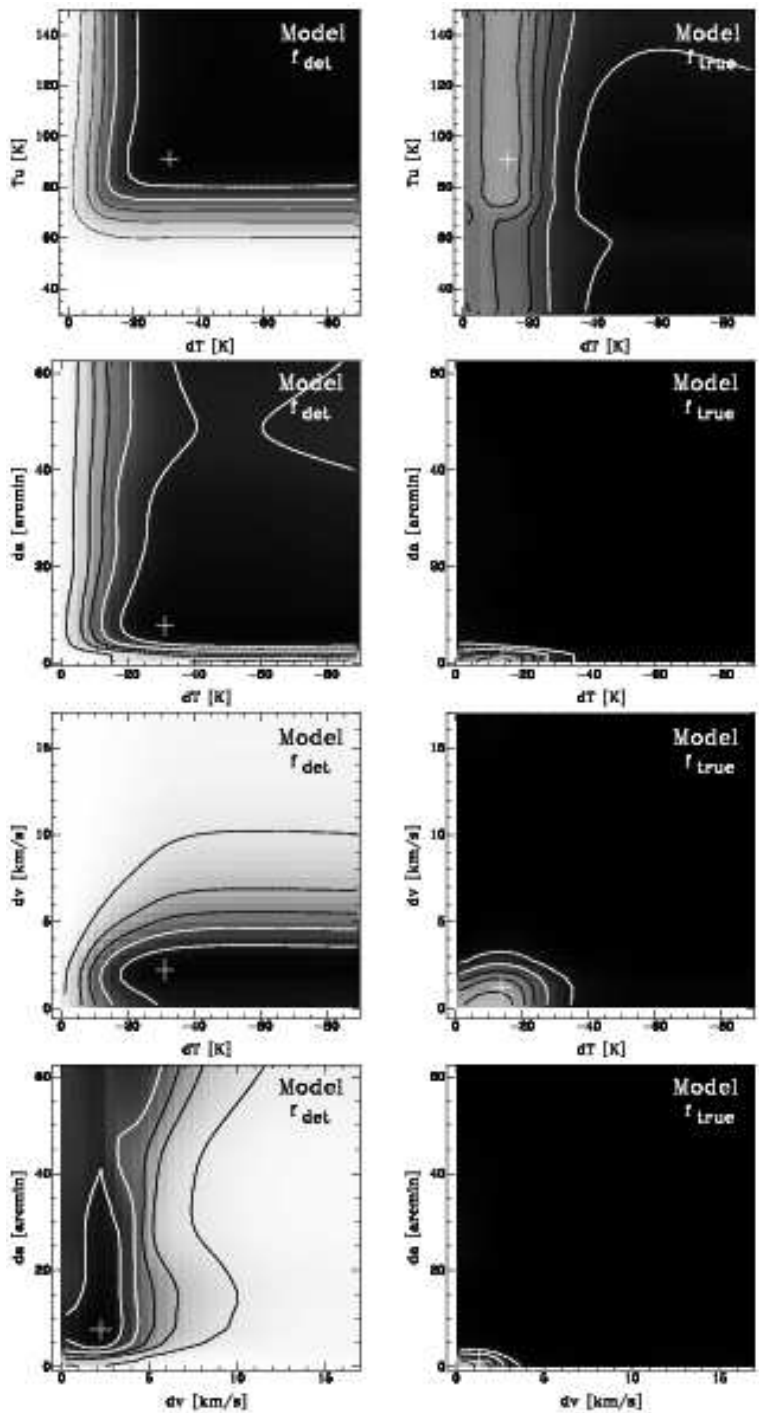}}}
\end{minipage}
\caption[]{
2-D slices through the 4-D throughput $\fid$ and true fraction $\fot$
histograms.  The $\fid$ slices intersect at a common position marked with a
cross.  The $\fot$ slices intersect at a different common position, also
marked.  The intensity scale is linear, from 0.0 (white) to 1.0 (black).
Black contours mark values of 0.1, 0.3, and 0.5; 
white contours mark values of 0.7 and 0.9.
Axis labels are as in Figure~\ref{Fig:4hist_counts}.
}
\label{Fig:4hist_fid_fot}
\end{figure}

Figure~\ref{Fig:4hist_fid_fot} presents selected 2-D slices through the 4-D
parameter space to illustrate the behavior of the throughput $\fid$ and true
fraction $\fot$.  These have been smoothed as described in
\S\ref{Sec:eval_anal_meas}.  We find that most HISA is detected if it is
significantly stronger than the noise, larger than a few beams, narrower than a
few \kms, and has $\tu \ga 80$~K.  Furthermore, the vast majority of detected
HISA is reliable, except for HISA that can be mimicked by beam-scale noise
fluctuations.

The throughput is high for much of the parameter space: $\fid \ga 0.80$ where
$\dt \la -20$~K, $\tu \ga 80$~K, $\da \ga 5$\arcmin, and $\dv \la 3.5$\kms,
reaching a maximum of $\sim 0.99$ where all of these criteria are well-met.
Where one or more of them is not met, $\fid$ drops rapidly, with $\fid
\rightarrow 0$ for $\dt \ga -2$~K, $\tu \la 60$~K, $\da \la 1$\arcmin, or $\dv
\ga 8$\kms.  Most of this behavior can be explained as losing features in the
noise, underbright $\tu$, or overbroad linewidths poorly fitted by the spectral
search method's $W_{broad}$ [4\kms\/] Gaussians (\S\ref{Sec:ext_spectral}).
However, low $\fid$ seems to occur for low $\da$ even when $|\dt|$ is large.
This suggests some beam-scale HISA features may be missed by our search if they
are isolated from larger structures.  The Perseus HISA Globule of Paper~I is
detected easily, but it is also attached to the complex GHISA
139.01$+$0.96$-$40.

By contrast, narrow-line HISA is detected with great efficiency: high $\fid$ is
found for $\dv$ as low as the 0.8\kms\ CGPS channel width, so long as $\dt \la
-10$~K.  To allay the concerns of \citet{lg03}, lines narrower than this should
also be detectable if their intrinsic amplitudes are larger to compensate for
spectral dilution.  A thermally broadened $\ts = 10$~K HISA line would be
diluted by a factor of 2.4 but easily detected if $\dt \la -24$~K.  HISA of
this strength or greater is common in the CGPS.

The true fraction has much simpler behavior: $\fot \rightarrow 1.0$ almost
everywhere that $\dt \la -20$~K, $\da \ga 3$\arcmin, or $\dv \ga 3$\kms.  If
none of these holds, noise fluctuations in the data produce significant false
positive detections, with $\fot \ga 0.2$ in the worst cases.

\subsubsection{Parameter Drift}
\label{Sec:eval_res_drift}

\begin{figure}
\begin{minipage}[t]{16cm}
\centerline{\resizebox{15cm}{!}{\includegraphics[angle=0]{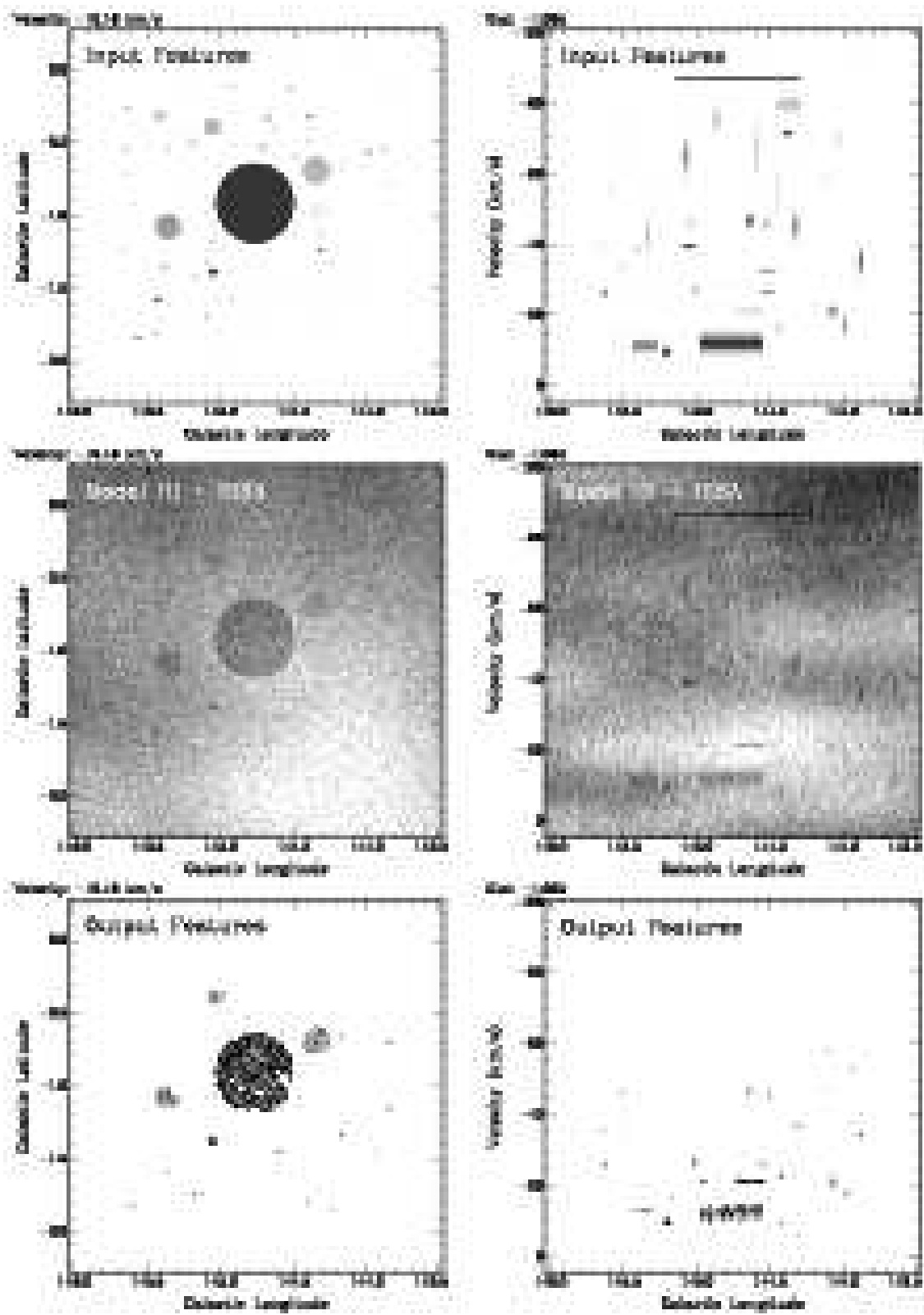}}}
\end{minipage}
\caption[]{
$(\glon,\glat)$ and $(\glon,\vel)$ slices of sample model \hi\ data, 
showing the input HISA ``hockey puck'' amplitude $\dti$, the noisy
background emission field $\tui$ with pucks added, and the extracted 
HISA amplitude $\dto$ after the 3rd identification pass.
Intensity ranges are $-40$ to 0~K for the $\dt$ maps and 0 to 120~K for the
\hi\ maps, from black to white.  
}
\label{Fig:hockey_pucks}
\end{figure}

\begin{figure}
\begin{minipage}[t]{16cm}
\centerline{\resizebox{15cm}{!}{\includegraphics[angle=0]{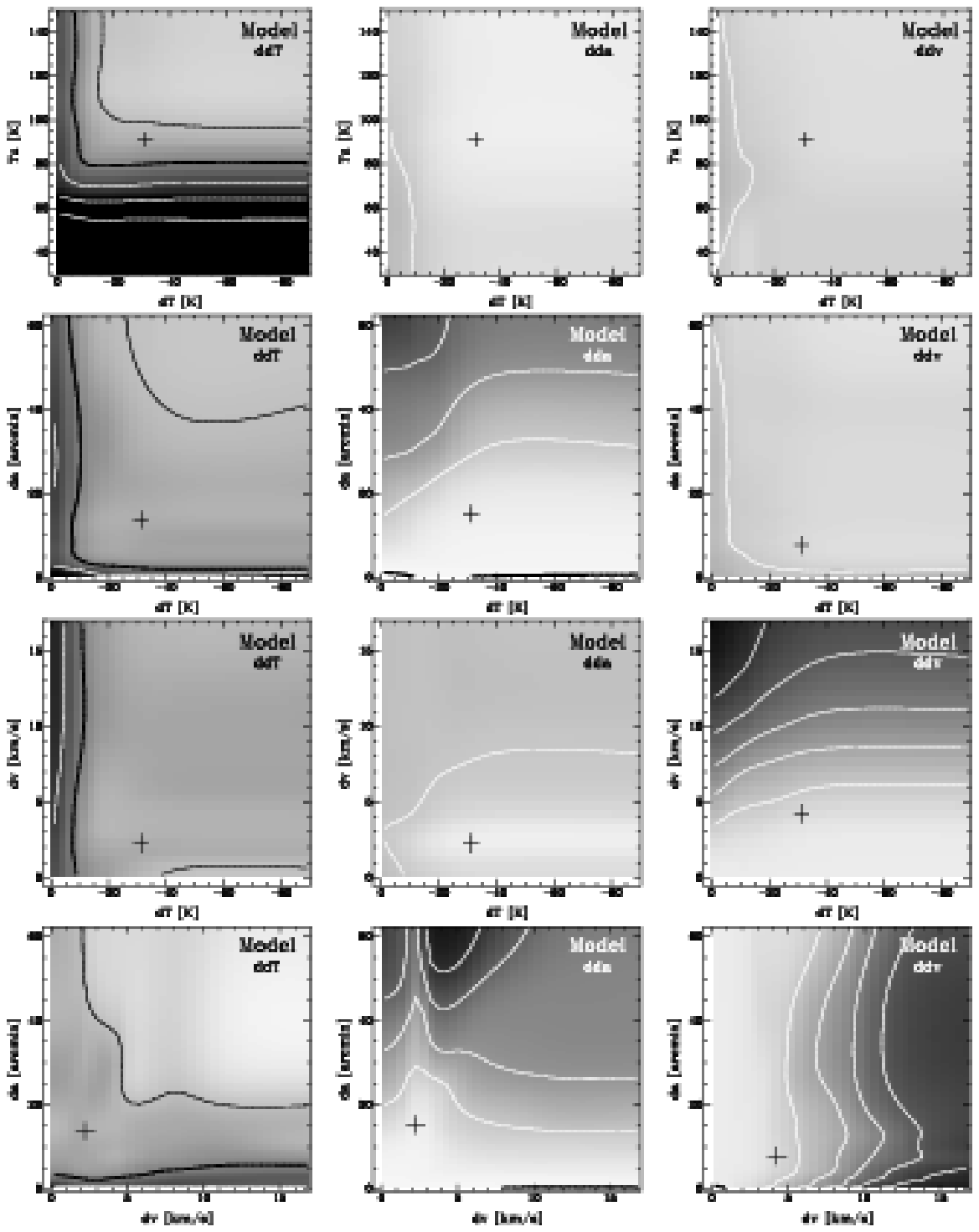}}}
\end{minipage}
\caption[]{
2-D slices through the 4-D drift histograms $\ddt$, $\dda$, and
$\ddv$.  As in Figure~\ref{Fig:4hist_fid_fot}, crosses mark 
slice intersections for each 4-D drift measure.  
The intensity scale is linear, from
negative (black) to
positive (white). 
The intensity ranges are 
$-10$~K $< \ddt < +10$~K, 
$-50\arcmin < \dda < +1\arcmin$, and 
$-13\kms < \ddv < +1\kms$.
Where present, a thick black contour marks zero drift.  Thinner 
contours mark positive (black) and negative (white) drifts at intervals
of 5~K, 10\arcmin, and 2\kms, respectively.  
Axis labels are as in Figure~\ref{Fig:4hist_counts}.
}
\label{Fig:4hist_drift}
\end{figure}

Figure~\ref{Fig:4hist_drift} illustrates trends in the parameter drifts $\ddt$,
$\dda$, and $\ddv$.  Since $\tu = \ton - \dt$ (\S\ref{Sec:ext_dt}) and $\ton$
is fixed, $\dtu = - \ddt$.  With minor exceptions, the behavior of all the
parameter drifts is fairly simple: $|\dt|$ and $\tu$ are often underestimated
by a few K in well-detected features, while incomplete detections of large
features (e.g., Fig.~\ref{Fig:hockey_pucks}) cause $\da$ and $\dv$ to be
underestimated as well.

The drift in $\dt$ is positive, i.e., toward reduced amplitudes, if these three
conditions are met: $|\dt| \ga$ the 6~K noise level, $\tu \ga$ 80~K, and $\da
\ga$ 2\arcmin.  If one of them is not met, $\ddt < 0$.  There is no strong
dependence on $\dv$.  The amount of drift is typically a few K, with a range of
$\pm 10$~K in most areas but more negative for $\tu < 65$~K.  The $\ddt$
behavior has a similar shape to $\fid$ above, suggesting that detection
sensitivity governs $\dt$ drift.  Features with intrinsically low $|\dt|$, low
$\tu$, or very small $\da$ appear to have larger $|\dt|$ (and $\tu$) if they
are detected.  But $\fid$ shows most are {\it not\/} detected; those that are
represent a biased sample in which $|\dt|$ and $\tu$ happened to be boosted in
the right direction to make them detectable.  By contrast, easily detectable
features appear to have lower $|\dt|$ than they should.  Since $\dtu = - \ddt$,
their $\tu$ is underestimated as well; one of the mechanisms noted in
\S\ref{Sec:eval_mot} may be to blame.  But whether positive or negative, the
magnitude of $\ddt$ is usually only a few K.  The large drifts that produced
the $\dt$ outliers in the model $\voa$ results (\S\ref{Sec:eval_res_dist};
Fig.~\ref{Fig:4hist_counts}) are exceptional cases.

The drift in $\da$ is negative everywhere that $\da >$ 1\arcmin.  It covers a
range of $-50\arcmin \la \dda \la +1\arcmin$, becoming steadily more negative
for larger $\da$, with only minor dependencies on other parameters.  The small
positive drifts occur when $\da < 1\arcmin$ HISA is augmented by beam-scale
noise fluctuations; $\da < 1\arcmin$ can occur in the line wings of 1\arcmin\
features, since pucks appear smaller off the line center
(\S\ref{Sec:eval_res_dist}).  The much larger negative drifts are from those
large features that are incompletely detected, as confirmed by visual
inspection of the $(\glon,\glat,\vel)$ data (e.g.,
Fig.~\ref{Fig:hockey_pucks}).  Some of these partial detections are caused by
$\tu \sim 70$~K boundaries, but most are from noise.  2- and 3-$\sigma$ noise
fluctuations frequently poke holes in fairly strong features, reducing their
$\da$ measures.  This is especially common in the line wings where $|\dt|$ is
less.

The drift in $\dv$ is positive only where both $\da < 2\arcmin$ and $\dv <
1\kms$ and is negative everywhere else.  It covers a range of $-13\kms < \dv <
+0.2\kms$, becoming steadily more negative for larger $\dv$, with only minor
dependencies on other parameters.  As with $\dda$, noise degradation is a major
cause of this $\ddv$ trend.  But in addition, the spectral HISA search itself
is optimized for the detection of HISA linewidths $\la 4\kms$, and Gaussian
features broader than 8\kms\ are CLEANed out almost entirely
(\S\ref{Sec:eval_res_fid_fot}).  Lastly, features near emission peaks may have
$\tu$ underestimated (\S\ref{Sec:eval_mot}), leading to $|\dt|$ underestimates
in line wings, and thus $\dv$ underestimation.  Since our $\tu$ estimation is
not a simple linear interpolation (\S\ref{Sec:ext_dt}), this effect may not be
as severe as that noted by \citet{lb80}, but the $\ddt > 0$ results above for
well-detected features suggest it is not zero either.

\citet{lb80} also note that $\tu$ gradients will cause $\dv$ to be narrower
than the HISA optical depth profile FWHM, and the line center in $\dt$ will
appear shifted from $\tau_{max}$, the maximum optical depth.  However, the
performance of our HISA software is only concerned with $\dt$, so these biases
do not apply here.  And while our voxel-based evaluation method is not able to
track changes of position, visual inspection shows that filaments and other
structures within features do not shift between input and output; only the
centroids of whole features may shift if the features are not completely
detected.

\subsection{Reliability and Completeness}
\label{Sec:eval_app}

\begin{figure}
\plotone{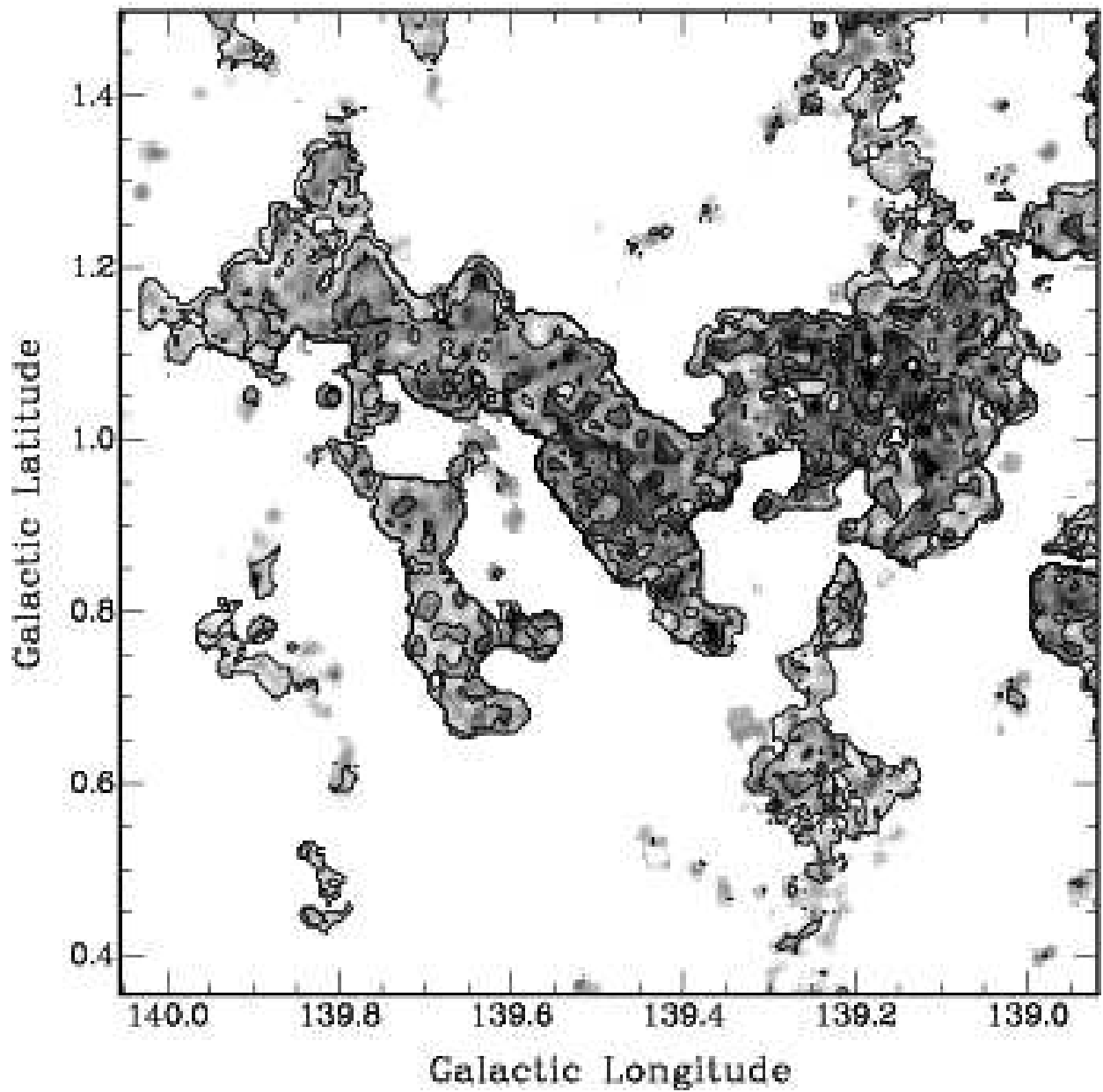}
\caption[]{
Sample extracted HISA $\dt(\glon,\glat)$ map for a single velocity, showing
contours of $\fot =$ 0.682690, 0.954500, 0.997300, and 0.999937,
which correspond to reliability threshholds of 
1, 2, 3, and 4 $\sigma$ if Gaussian statistics apply.
The maximum $\fot$ in this map is 0.999999, which is equivalent to 4.90 
$\sigma$.  The region shown is the same as in Figure~\ref{Fig:da_dv_examples}.
}
\label{Fig:feature_with_fot}
\end{figure}

These results have many uses.  In addition to statistically describing how well
HISA is detected, they can be applied directly to particular features to assess
their reliability and completeness.  The former is done by measuring
$(\dt,\tu,\da,\dv)$ at each HISA voxel $(\glon,\glat,\vel)$ position and
interpolating $\fot$ from the 4-D histogram described in
\S\ref{Sec:eval_res_fid_fot}.  $\fot$ is the likelihood that a HISA detection
represents real absorption.  We have determined $\fot$ for each CGPS HISA
voxel.  Figure~\ref{Fig:feature_with_fot} shows $\fot$ contours on a sample
HISA feature.  The HISA detection reliability in this case is quite high.  In
Paper~II, we use $\fot(\glon,\glat,\vel)$ in analyses of total CGPS HISA
coverage and the distributions of weak and strong absorption.  We also assess
the completeness of our HISA detections.  The actual detection fraction $\fid$
is not recoverable from observed data, but we consider the fraction of
detections with $\langle \tu \rangle > 80$~K, since such HISA has $\fid > 0.8$
if its size and strength are appreciable (\S\ref{Sec:eval_res_fid_fot}).

\section{Conclusions}
\label{Sec:conclusions}

We have described algorithms that identify and extract \hi\ self-absorption
(HISA) features in high-resolution \hi\ 21cm line data cubes.  These algorithms
were designed to carry out a HISA survey of cold \hi\ in the initial $73\arcdeg
\times 9\arcdeg$ phase of the arcminute-resolution Canadian Galactic Plane
Survey (CGPS), but they should have more general applicability.

Our search algorithms use CLEAN-based spatial and spectral filtering to remove
large-scale emission structure and identify HISA as significant negative
residuals.  Features identified in both spectral and spatial domains are
flagged as HISA, and the unabsorbed brightness $\tu$ along the feature
sightline is estimated from a 3-D interpolation of the OFF-feature brightness
temperature $\toff$.  HISA detections in overly noisy regions are rejected, as
are those for which $\tu < 70$~K, lest significant false detections result from
gaps between sharply-structured emission features with faint backgrounds.  In
order to capture features larger than the CLEAN filter scale, identified HISA
is removed and the search process is repeated; a total of three such passes
suffices for the CGPS data.

We performed detailed tests of our HISA-finding software with model data to
determine its detection limits, false positive rates, and measurement biases as
functions of feature size, amplitude, and background field brightness.  The
tests show that HISA is well detected within the software design criteria, with
high detection rates for HISA significantly stronger than the noise level,
larger than a few beams, narrower than a few \kms, and with $\tu \ga 80$~K.  At
the same time, the bulk of HISA detections are reliable, with very low false
positive rates in most parts of the parameter space except those occupied by
beam-scale noise fluctuations.  Measurement drifts are small in well-detected
features, with $\tu$ underestimated by a few K due to contamination of $\toff$
by faint, undetected HISA near the feature.  Where detections are truncated by
noise fluctuations or faint $\tu$, the bias may be somewhat larger.  Incomplete
detections also make features appear smaller in angular size and linewidth than
in reality due to truncation.

This paper is the third in an ongoing series investigating HISA at high
resolution in the Galactic plane.  A companion paper (Paper~II) presents HISA
survey results for the CGPS.  Subsequent papers will further analyze the CGPS
HISA and also examine HISA in CGPS extensions and in the VLA Galactic Plane
Survey.

\acknowledgements

We thank W. McCutcheon, T. Landecker, and J. Stil for a number of useful
discussions on this project, and the anonymous referee for constructive
comments on the manuscript.  J. Stil helped with multipanel figure layout.  We
are very grateful to R. Gooch for tireless computing support, including
continued expansion of the capabilities of the Karma visualization software
package \citep{karma96}\footnote{See also
\url{http://www.atnf.csiro.au/karma}.}, which was used extensively for this
work.  The Dominion Radio Astrophysical Observatory is operated as a national
facility by the National Research Council of Canada.  The Canadian Galactic
Plane Survey (CGPS) is a Canadian project with international partners.  The
CGPS is described in \citet{cgps}\footnote{Additional information is available
online at \url{http://www.ras.ucalgary.ca/CGPS}.}.  The main CGPS data set is
available at the Canadian Astronomy Data Centre\footnote{See
\url{http://cadcwww.hia.nrc.ca/cgps}.}.  The CGPS is supported by a grant from
the Natural Sciences and Engineering Research Council of Canada.

\appendix
\section{Corrections to Paper I Results}
\label{App:erratum}

\begin{figure}[ht]
\plotone{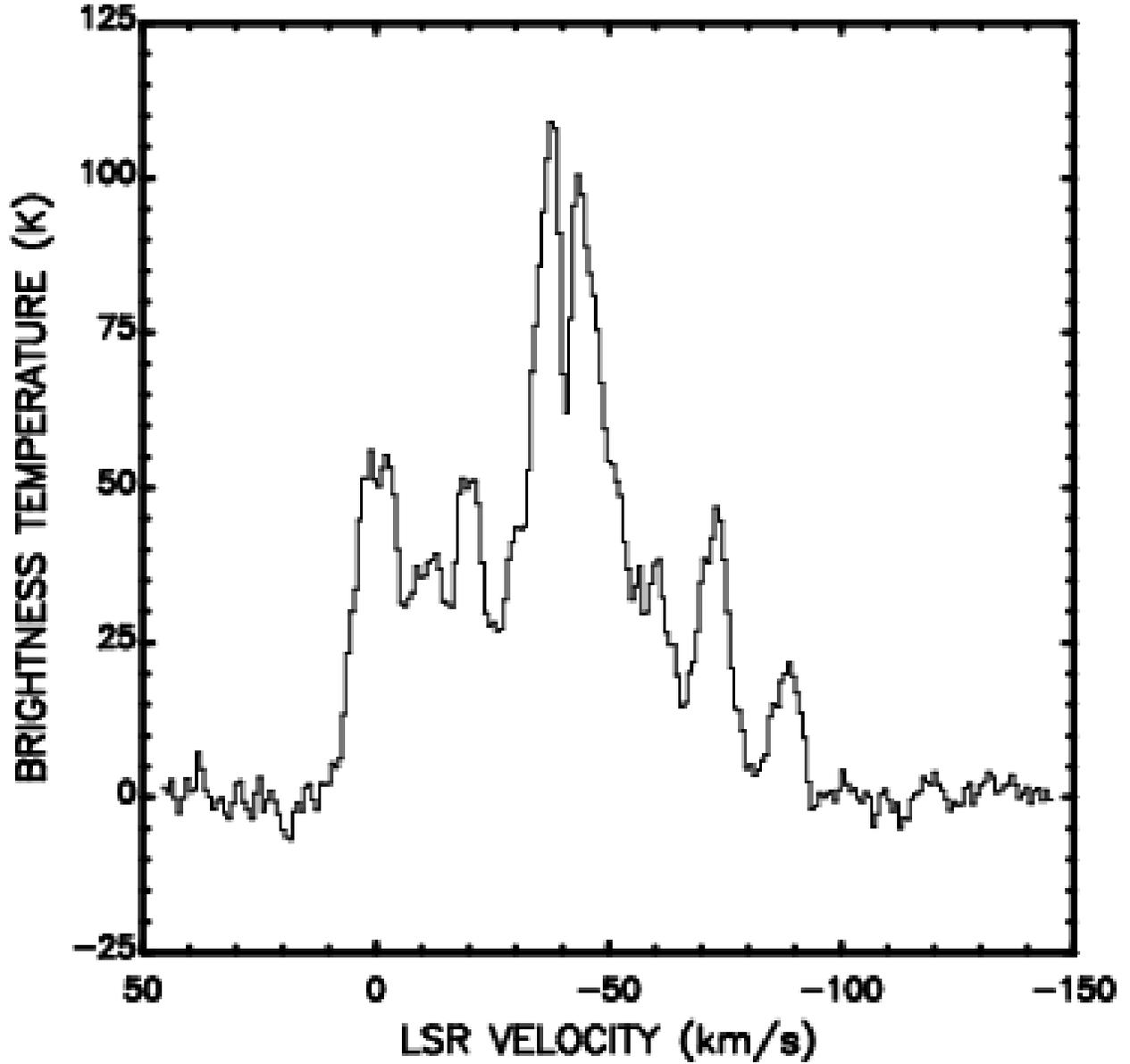}
\caption {
The full-resolution observed spectrum $O(k)$ at the 
($\glon = 139.635\arcdeg, \, \glat = 1.185\arcdeg$) position of the Perseus 
HISA Globule of Paper~I.  The feature's absorption 
amplitude has changed from Paper~I due to correction of 
a data processing error that had no serious impact on the derived results
(see Appendix~\ref{App:erratum}).
}
\label{Fig:globule_spectrum_raw}
\end{figure}

\begin{figure}
\begin{minipage}[t]{16cm}
\centerline{\resizebox{15cm}{!}{\includegraphics[angle=0]{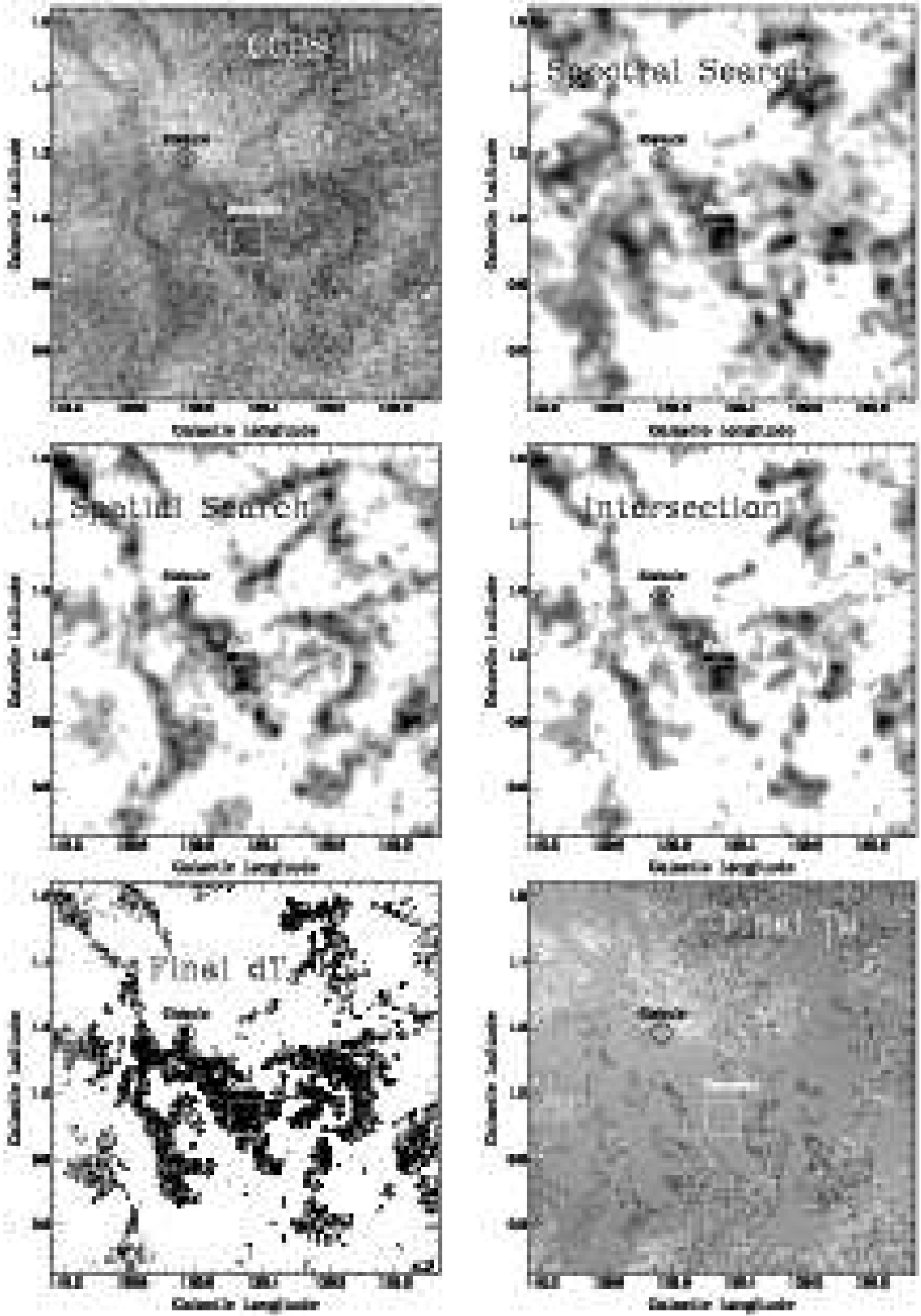}}}
\end{minipage}
\caption[]{
Detailed views of each panel in Figure~\ref{Fig:real_hisa_id_maps_overview}, 
showing the same area as Figure~2 of Paper~I.  
The Perseus HISA Complex and Globule positions of that paper are marked.  
Sample velocity spectra of the Globule are given in 
Figure~\ref{Fig:real_hisa_id_spectra}.
}
\label{Fig:real_hisa_id_maps_zoom}
\end{figure}

\begin{figure}
\plotone{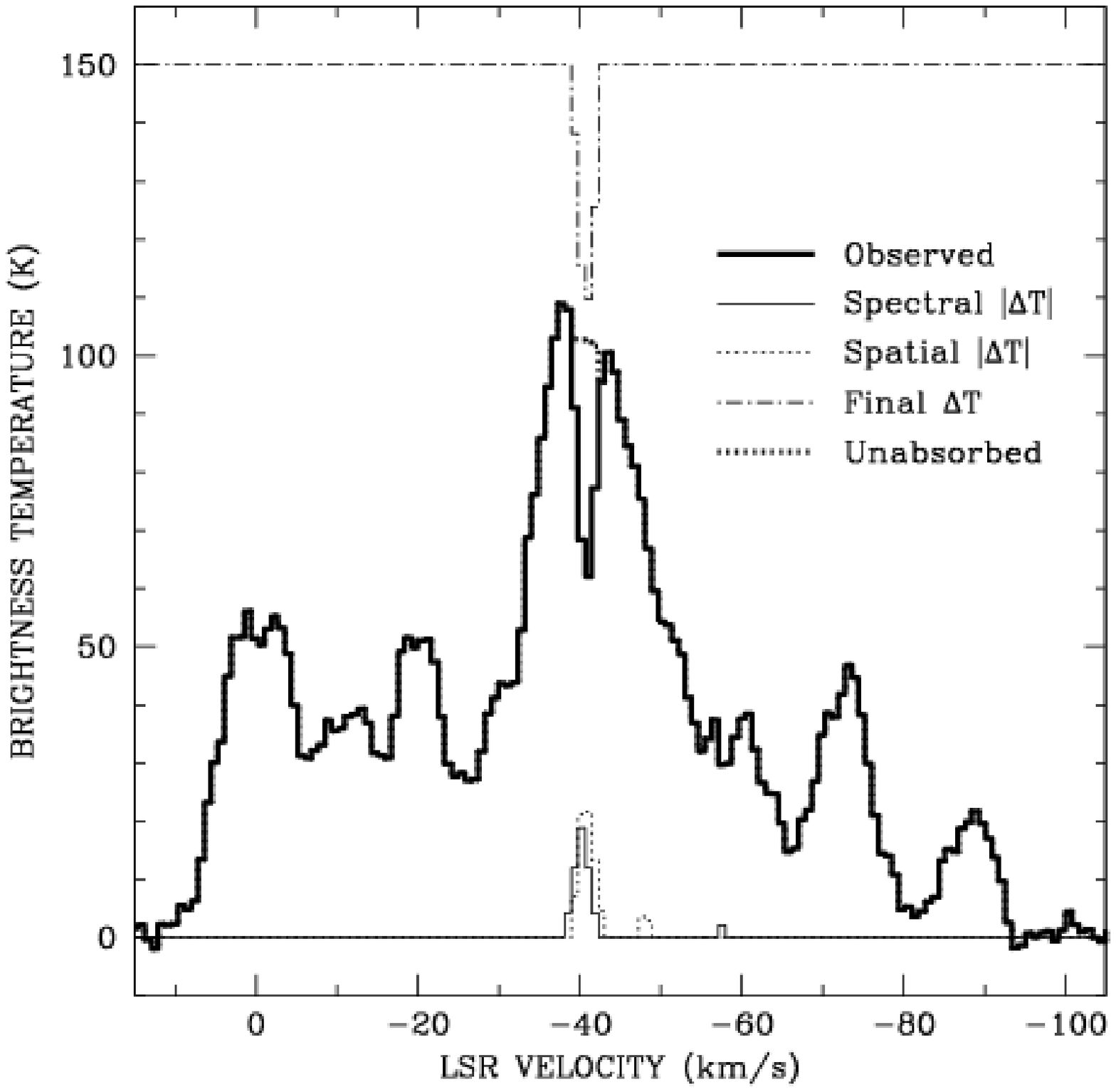}
\caption[]{
Single-pixel velocity spectra at the Perseus HISA Globule position
($\glon = 139.635\arcdeg, \, \glat = 1.185\arcdeg$), showing
CGPS \hi, HISA $|\dt|$ from the 
spectral (\S\ref{Sec:ext_spectral}) and 
spatial searches (\S\ref{Sec:ext_spatial}), 
and the final $\dt$ and $\tu$ from the full 3-D extraction
(\S\ref{Sec:ext_dt}); the final $\dt$ zero point has been shifted to 150~K
for clarity.  
Although $\tu$ appears below the observed $\toff$ at the HISA velocity edges, 
this spectrum shows only a small subset of the $\toff$ voxels 
that surround the feature in 3-D.  $\tu$ is estimated from the entire 3-D 
$\toff$ set, a larger sample of which is
shown in the corresponding channel map in 
Figure~\ref{Fig:real_hisa_id_maps_zoom}.
}
\label{Fig:real_hisa_id_spectra}
\end{figure}

The \hi\ data in the vicinity of $(\glon = 140\arcdeg,\glat=+1\arcdeg)$ have
been revised from those used in Paper~I.  A single synthesis field was assigned
the wrong flux scale in the \hi\ data used in that paper, and this error was
not discovered until after publication.  As a result, the HISA amplitudes
presented in Paper~I for the Perseus HISA Complex and Globule features were in
error, and the correct HISA amplitudes are smaller than those found in Paper~I.
With revised data, these features have warmer spin temperatures and lower
optical depths than those derived in Paper~I, but the column densities and
masses are only mildly affected.  Table~\ref{Tab:cg_prop} lists the corrected
results for both features.  The correct Globule spectrum is plotted in
Figure~\ref{Fig:globule_spectrum_raw}, and the positions of both features are
marked in Figure~\ref{Fig:real_hisa_id_maps_zoom}.  The Local HISA Filament
presented in Paper~I was unaffected by this problem.

\clearpage

\begin{deluxetable}{ll|cc|cc}
\tablewidth{0pt}  
\tablecaption{Corrected Perseus HISA Complex and Globule Properties\tablenotemark{*}}
\tablehead{
 \multicolumn{2}{c}{} & \multicolumn{2}{c}{Perseus} & \multicolumn{2}{c}{Perseus} \\
 \multicolumn{2}{c}{} & \multicolumn{2}{c}{Complex} & \multicolumn{2}{c}{Globule} \\
 \multicolumn{2}{r}{\hi\ Data:} & \colhead{Paper I} & \colhead{Revised} & \colhead{Paper I} & \colhead{Revised}
}
\tablecolumns{6}
\startdata
\multicolumn{6}{l}{Input Parameters\tablenotemark{\dag}} \\
\hline
& $\ton$ [K]       & ~~69 & ~~71 & ~47 & ~~62 \\
& $\toff$ [K]      & ~107 & ~~99 & 112 & ~104 \\
& $\dt$ [K]        & $-38$ & $-28$ & $-65$ & $-42$ \\
\hline
\multicolumn{4}{l}{Derived Gas Properties $(\bfrac = \hifrac = 1)$} & \\
\hline
& $\ts$ [K] & $45-61$ & $49-65$ & $32-35$ & $41-43$ \\
& $\tau$ & $0.83-1.37$ & $0.71-1.22$ & $1.43-1.54$ & $0.94-0.99$ \\
& $\coldens$ [10$^{20}$~cm$^{-2}$] & $3.2-7.3$ & $3.0-6.8$ & $2.2-2.6$ & $1.8-2.0$ \\
& $\voldens$ [cm$^{-3}$] & $89-65$ & $81-62$ & $124-115$ & $99-94$ \\
& $\mass$ [\msun] & $31-111$ & $32-106$ & $0.60-1.09$ & $0.53-0.80$ \\
\hline
\multicolumn{6}{l}{Derived Gas Properties ($\hifrac = 0.01$, Maximum Total Mass)} \\
\hline
& $\ts$  [K] & $2.7$ & $2.7$ & $2.7$ & $2.7$ \\
& $\tau$ & $7.0$ & $6.9$ & $2.5$ & $2.4$ \\
& $\coldens$    [10$^{20}$~cm$^{-2}$] & $1.7$ & $1.6$ & $0.33$ & $0.32$ \\
& $\voldens$    [cm$^{-3}$] & $15$ & $15$ & $15$ & $15$ \\
& $\mass$       [\msun] & $26$ & $25$ & $0.14$ & $0.12$ \\
& $N_{tot}$    [10$^{20}$~cm$^{-2}$] & $170$ & $160$ & $33$ & $32$ \\
& $n_{tot}$    [cm$^{-3}$] & $1500$ & $1500$ & $1500$ & $1500$ \\
& $M_{tot}$ [\msun] & $5200$ & $5000$ & $28$ & $25$ \\
\enddata
\label{Tab:cg_prop}
\tablenotetext{*}{This table follows the format of Table 1 in Paper I.}
\tablenotetext{\dag}{Only input parameters that have changed from Paper I are shown.}
\end{deluxetable} 

\clearpage

\clearpage

\clearpage

\clearpage

\clearpage

\clearpage

\begin{thebibliography}{}
\bibitem[Baker \& Burton(1979)]{bb79} Baker, P. L., \& Burton, W. B. 1979, 
  \aaps, 35, 129
\bibitem[Bania \& Lockman(1984)]{bl84} Bania, T. M., \& Lockman, F. J. 1984, 
  \apjs, 54, 513
\bibitem[Colgan et al.(1988)Colgan, Salpeter, \& Terzian]{cst88}
  Colgan, S. W. J., Salpeter, E. E., \& Terzian, Y. 1988, \apj, 328, 275
\bibitem[Feldt(1993)]{f93} Feldt, C. 1993, \aap, 276, 531
\bibitem[Gibson(2002)]{g02} Gibson, S. J. 2002, ASP Conf.\ Ser.\ 276, {\it 
  Seeing Through the Dust: the Detection of \hi\ and the Exploration of the
  ISM in Galaxies}, eds.\ A. R. Taylor, T. L. Landecker, \& A. G. Willis, 235
\bibitem[Gibson et al.(2000)]{g00} Gibson, S. J., Taylor, A. R., 
  Dewdney, P. E., \& Higgs, L. A. 2000, \apj, 540, 851 (Paper~I)
\bibitem[Gibson et al.(2005)]{g05a} Gibson, S. J., Taylor, A. R., Higgs, L.
  A., Brunt, C. M., \& Dewdney, P. E. 2005, \apj, submitted (Paper~II)
\bibitem[Gibson et al.(2004)]{g04} Gibson, S. J., Taylor, A. R., Stil, J. M.,
  Higgs, L. A., Dewdney, P. E., \& Brunt, C. M. 2004, in {\it How Does the
  Galaxy Work?  A Galactic Tertulia with Don Cox and Ron Reynolds}, eds.\ 
  E. J. Alfaro, E. P\'{e}rez, \& J. Franco (Dordrecht: Kluwer Academic
  Publishers), 47
\bibitem[Gooch(1996)]{karma96} Gooch, R. 1996, ASP Conf.\ Ser.\ 101, {\it 
  Astronomical Data Analysis Software and Systems V}, eds. G. H. Jacoby \& 
  J. Barnes, 80
\bibitem[Green(1993)]{g93} Green, D. A. 1993, \mnras, 262, 327
\bibitem[Hasegawa et al.(1983)Hasegawa, Sato, \& Fukui]{hsf83} Hasegawa, T., 
  Sato, F., \& Fukui, Y.  1983, \aj, 88, 658
\bibitem[H\"{o}gbom(1974)]{h74} H\"{o}gbom, J. A. 1974, \aaps, 15, 417
\bibitem[Kavars et al.(2003)]{k03} Kavars, D. W., Dickey, J. M., 
  McClure-Griffiths, N. M., Gaensler, B. M., \& Green, A. J. 
  2003, \apj, 598, 1048
\bibitem[Kerton(2005)]{k05} Kerton, C. R. 2005, \apj, accepted
\bibitem[Knapp(1974)]{k74} Knapp, G. R. 1974, \aj, 79, 527
\bibitem[Knee \& Brunt(2001)]{kb01} Knee, L. B. G., \& Brunt, C. M. 2001, \nat,
  412, 308
\bibitem[Levinson \& Brown(1980)]{lb80} Levinson, F. H., \& Brown, R. L.
  1980, \apj, 242, 416
\bibitem[Li \& Goldsmith(2003)]{lg03} Li, D., \& Goldsmith, P. F. 2003, \apj,
  585, 823
\bibitem[McClure-Griffiths et al.(2001)]{sgps} McClure-Griffiths, N. M.,
  Green, A. J., Dickey, J. M., Gaensler, B. M., Haynes, R. F., \& Wieringa, M.
  H. 2001, \apj, 551, 394.
\bibitem[McCutcheon et al.(1978)McCutcheon, Shuter, \& Booth]{msb78} 
  McCutcheon, W. H., Shuter, W. L. H., \& Booth, R. S. 1978, \mnras, 185, 755
\bibitem[Minter et al.(2001)]{m01} Minter, A. H., Lockman, F. J., Langston, 
  G. I., \& Lockman, J. A. 2001, \apj, 555, 868
\bibitem[Montgomery et al.(1995)Montgomery, Bates, \& Davies]{mbd95} 
  Montgomery, A. S., Bates, B., \& Davies, R. D. 1995, \mnras, 273, 449
\bibitem[Peebles(1993)]{p93} Peebles, P. J. E. 1993, {\it Principles of 
  Physical Cosmology} (Princeton: Princeton University Press), 35
\bibitem[Peters \& Bash(1987)]{pb87} Peters, W. L., \& Bash, F. N. 1987, \apj,
  317, 646
\bibitem[Riegel \& Crutcher(1972)]{rc72} Riegel, K. W., \& Crutcher, R. M.
  1972, \aap, 18, 55
\bibitem[Steer et al.(1984)Steer, Dewdney, \& Ito]{sdi84} Steer, D. G., 
  Dewdney, P. E., \& Ito, M. R. 1984, \aap, 137, 159
\bibitem[Taylor et al.(2003)]{cgps} Taylor, A. R., et al.\ 2003, \aj, 125, 
  3145
\bibitem[Taylor et al.(2002)]{vgps} Taylor, A. R., Stil, J. M., Dickey, J. M.,
  McClure-Griffiths, N. M., Martin, P. G., Rothwell, T., \& Lockman, F. J. 
  2002, ASP Conf. Ser. 276, {\it Seeing Through the Dust: the Detection of
  \hi\ and the Exploration of the ISM in Galaxies}, eds. A. R. Taylor, T. L.
  Landecker, \& A. G. Willis, 68
\bibitem[van der Werf et al.(1988)van der Werf, Goss, \& Vanden Bout]{vdw88} 
  van der Werf, P. P., Goss, W. M., \& Vanden Bout, P. A. 1988, \aap, 201, 311
\end{thebibliography}
\end{document}